\documentclass[preprint,journal]{vgtc}       

\ifpdf
  \pdfoutput=1\relax                   
  \usepackage{graphicx}                
  \DeclareGraphicsExtensions{.pdf,.png,.jpg,.jpeg} 
\else
  \usepackage{graphicx}                
  \DeclareGraphicsExtensions{.eps}     
\fi%

\graphicspath{{figures/}{pictures/}{images/}{./}} 

\usepackage{microtype}                 
\PassOptionsToPackage{warn}{textcomp}
\usepackage{textcomp}
\usepackage{mathptmx}
\usepackage{times}

\usepackage{cite}
\usepackage{tabu}
\usepackage{booktabs}

\usepackage{hyperref}
\usepackage{url}
\usepackage{xcolor}

\usepackage{graphicx}
\usepackage{epstopdf}
\usepackage{booktabs}
\usepackage{float}
\usepackage{caption}

\usepackage{amsmath}
\usepackage{bm}
\usepackage{algorithm}
\usepackage{algpseudocode}
\usepackage{amsmath}
\usepackage{multirow}
\usepackage{indentfirst}
\usepackage{subfig}
\usepackage{epsfig}
\usepackage{epstopdf}
\usepackage{xspace}

\usepackage{subeqnarray}

\usepackage{color}
\usepackage[utf8]{inputenc}
\usepackage{siunitx}
\usepackage{rotating}

\usepackage{paralist}
\usepackage{cleveref}

\usepackage{multicol}
\title{Comparative Visual Analytics for Assessing Medical Records with Sequence Embedding}

\author{Rongchen Guo\thanks{\small{Corresponding author. E-mail: rongchen.guo1020@gmail.com}} \thanks{\small{Department of Computer Science, Beihang University}}~, Takanori Fujiwara\thanks{\small{Department of Computer Science, University of California, Davis}}~, Yiran Li\footnotemark[3]~, Kelly~M.~Lima\thanks{\small{Department of Pathology and Laboratory Medicine, University of California, Davis}}~, Soman Sen\thanks{\small{Department of Surgery, University of California, Davis}}~, Nam~K.~Tran\footnotemark[4]~, and Kwan-Liu Ma\footnotemark[3]}

\abstract{
Machine learning for data-driven diagnosis has been actively studied in medicine to provide better healthcare.  Supporting analysis of a patient cohort similar to a patient under treatment is a key task for clinicians to make decisions with high confidence. However, such analysis is not straightforward due to the characteristics of medical records: high dimensionality, irregularity in time, and sparsity. To address this challenge, we introduce a method for similarity calculation of medical records. Our method employs event and sequence embeddings. While we use an autoencoder for the event embedding, we apply its variant with the self-attention mechanism for the sequence embedding. Moreover, in order to better handle the irregularity of data, we enhance the self-attention mechanism with consideration of different time intervals. We have developed a visual analytics system to support comparative studies of patient records. To make a comparison of sequences with different lengths easier, our system incorporates a sequence alignment method. Through its interactive interface, the user can quickly identify patients of interest and conveniently review both the temporal and multivariate aspects of the patient records. We demonstrate the effectiveness of our design and system with case studies using a real-world dataset from the neonatal intensive care unit of UC Davis.
}


\vgtcinsertpkg

\begin{document}

\firstsection{Introduction}
\maketitle

\label{sec:introduction}

Data-driven clinical decision-making using patient health information stored as Electronic Medical Records (EMRs) has been an area of research as a pathway towards precision medicine~\cite{yazhini2019state,harerimana2019deep}.
EMRs contain time sequence data recording patient visits to hospitals. These data include patient medical history and diagnoses, lab tests, medications, procedures, among others~\cite{lee2017big}.
Analyzing EMR data with recent advanced machine learning methods, specifically deep learning (DL) methods, gives promise to improved accuracy and precision of clinical decision making.
For example, DL-based analysis may identify future risks of a surgical procedure~\cite{kawaler2012learning}, deterioration of the patients' health condition during the treatment periods~\cite{mortalitypred}, and many more~\cite{pham2017predicting,huang2019patient}. 

In spite of the potential benefits above, many of the DL-based methods on EMRs only show predicted results (e.g., prediction of heart failure~\cite{nguyen2016mathtt}, hospital re-admission~\cite{yadav2016deep}, and the best diagnosis allocation~\cite{li2019intelligent}), and they are often black-box models (i.e., difficult to understand why the model makes a particular prediction). 
When diagnosing and treating a patient, accessing to information about previous patients who had similar medical conditions to the current patient would help the clinicians make more confident decisions along with their domain knowledge and past experiences~\cite{jin2019carepre}. 
Also, reviewing similar medical records helps clinicians gain a deeper understanding of the progression patterns of a certain disease~\cite{zhu2016measuring}.

However, identifying and analyzing similar medical records is a challenging task due to the characteristics of EMR data. 
EMRs are often high-dimensional (e.g., many different medical tests),  irregular (i.e., patients have different data collection time points and intervals according to when they visit hospitals), and sparse (e.g., some patients take only specific medical tests among many others)~\cite{lee2017big}. 
Therefore, to identify similar medical records, we are in need of a similarity calculation method that can handle all of these characteristics. 
Also, as for the analysis, comparison of such complex data becomes a crucial task.

To address the challenges of handling high-dimensionality, irregularity, and sparsity, first, we have developed a method for identifying similar medical records with the focal patient by coupling and extending two existing embedding methods. 
We employ an autoencoder~\cite{kramer1991nonlinear} based event-embedding in order to handle many different event types (e.g., each patient's medical test results at each hospital visit) as more compact vector representations.
Then, to convert records with different lengths (e.g., different numbers of hospital visits) to vectors of the same length, we use  sequence to sequence learning (seq2seq)~\cite{sutskever2014sequence}, which uses the multilayered Long Short-Term Memory (LSTM) model~\cite{hochreiter1997long}. 
Additionally, when using seq2seq, our method extends the standard self-attention mechanism~\cite{lin2017structured} as sequence temporal self-attention to model irregular temporal information of medical records. 
Through these embedding steps, each patient's medical records can be represented as a single vector that has the same length as other patients, making possible  
computing the similarities of the patients' records.

We have also developed an interactive visual analytics system to facilitate 
comparative study of similar medical records.
The system helps clinicians effectively compare records with different lengths by employing a dynamic time warping~\cite{DTW} based alignment.
Also, the system provides overviews that are designed to help find salient changes in clinical progressions.
Additionally, to support finding patients who have taken similar tests from a sparse dataset, we visualize this similarity information with dimensionality reduction and clustering methods.
Using these functionalities with linked views showing detailed information, clinicians can understand how and why patients are similar or which aspects are different even though they are generally similar. 
Lastly, we demonstrate the effectiveness of our design and system with case studies using a dataset obtained from the Neonatal Intensive Care Unit of UC Davis.

\section{Related Work}

In this section, we survey the relevant works in (1) visual analytics of healthcare data and similar data records and (2) methods to measure the similarity of medical records.

\subsection{Visual Analytics of Healthcare and Similar Data Records}

Recognizing the growing availability of healthcare data from EMRs and Electronic Health Records (EHRs), researchers have developed various visual analytics methods.
These methods have focused on a wide range of topics, such as clinical decision support systems (CDSS)~\cite{jin2019carepre,Berner2016CDSSoverview}, interpretable machine learning for medical decisions~\cite{kwon2018retainvis,ji2019visual,InterpretProto2019,li2017deep}, and exploration of disease internal progressions~\cite{guo2018visual,kwon2019dpvis}. 

Among the works above, temporal event-sequence visualizations
of healthcare data closely relate to our work.
These works focus on revealing the frequent patterns of disease progression by summarizing EMR/EHR data into flow-based representations~\cite{perer2015mining,monroe2013temporal}. 
Furthermore, to clearly display a sequence pattern, Guo et al.~\cite{eventThread,guo2018visual} segmented sequences into latent stages to infer the disease progression.
Also, both methods by Monroe et al.~\cite{monroe2013temporal} and Gotz et al.~\cite{gotz2014decisionflow} highlight the key events to reduce visual complexity by filtering out unimportant events.
These works can help clinicians understand the transition patterns of a certain disease. 
Instead of analyzing the temporal event sequence as it is, other works have tried to extract meaningful information from the sequence. 
For example, OutFlow~\cite{wongsuphasawat2011outflow,wongsuphasawat2012exploring} bridges a connection between the sequence and its subsequent outcome (e.g., death or survival; a better or worse condition) of the corresponding patient.
CarePre~\cite{jin2019carepre} predicts the next medical events (e.g., potential diseases a patient may have) from the sequence and provides interactive visualizations to help understand the prediction with the patient's historical medical records. 
Also, similar to ours, to fulfill the clinicians' practical demands listed up through the interview, CarePre supports a comparison of medical records between one selected patient and similar patients.

When including applications that focus on a more general domain, 
there are several works visualizing similar historical records of people to help them make decisions in their lives (e.g., which actions Ph.D.~students should take to be a professor~\cite{du2016eventaction}). 
For example, from multiple sets of records, EventAction~\cite{du2016eventaction} finds similar sub-sequences by using a fixed size of the sliding window. 
Then, EventAction provides a functionality of comparing the current user's records with the sub-sequences related to the desired outcome (e.g., becoming a professor) with what-if analysis.
In PeerFinder~\cite{du2017finding} and LikeMeDonuts~\cite{du2019visual}, instead of fully automatic selection of similar records, Du et al.~\cite{du2017finding, du2019visual} provided the selection of similarity criteria which can be interactively changed, and they visualized similar records with tables~\cite{du2017finding} or sunburst diagrams~\cite{du2019visual}. 

Among the existing works above, the works in \cite{du2016eventaction,du2017finding,du2019visual,jin2019carepre} are most closely related to our work in terms of providing visual analytics methods for identifying similar records.
However, the works~\cite{du2016eventaction,du2017finding,du2019visual} did not specifically target medical records and their similarity calculations do not consider the data characteristics of EMRs (i.e., irregularity and sparsity). 
On the other hand, while CarePre~\cite{jin2019carepre} is designed for medical records, their similarity calculation uses dynamic-time warping (DTW). As described in \autoref{sec:model}, using DTW has a limited capacity to handle the irregularity. 
Another major difference is that our visual analytics system provides effective functionalities focusing on analyzing similar medical records (refer to \autoref{sec:vis}). 

\subsection{Similarity Calculation for Medical Records}
\label{sec:relatedmodel}

Here, we describe methods which can be used for calculating the similarity of temporal sequences with different length, including medical records.
These methods can be categorized into two different types: elastic measures using dynamic-programming and methods utilizing machine learning (ML).

The methods in the first category find the best alignment between two different sequences and calculate the similarity between sequences by computing a certain aggregation distance between the aligned event pairs. 
Dynamic time wrapping (DTW)~\cite{DTW}, longest common subsequence (LCSS)~\cite{LCSS}, edit distance with real penalty (ERP)~\cite{ERP}, and edit distance on real sequences (EDR)~\cite{EDR} are in this category. 
DTW aligns two temporal sequences with the best matching and uses the sum-of-pairs distance for the aligned series to denote sequence similarity. 
As DTW is vulnerable to noise events, (i.e., noise may lead to a big distance between two similar sequences), the LCSS and EDR are adopted to address this issue by skipping noise events. 
While DTW, LCSS, and EDR are non-metric distance functions, ERP aims to provide a metric distance function by introducing a constant reference event for  similarity calculation. 
However, as discussed about DTW in \cite{ratanamahatana2005three}, the methods above do not show strong advantages in handling sequences of different lengths.

To address the above problem, ML-based methods have been developed.
Most ML-based methods utilize seq2seq~\cite{sutskever2014sequence} as their basis~\cite{li2018deep}. 
seq2seq consists of two major parts, namely an encoder and a decoder. 
The encoder embeds an input sequence ${s}$ to a fixed-sized latent vector ${v}$ using one recurrent neural network (RNN)~\cite{rumelhart1986learning} and then the decoder, using another RNN, generates the target time series ${\widehat{s}}$ based on the vector ${v}$. 
seq2seq obtains the best representation (i.e., the latent vector ${v}$) by minimizing the error between the inputs ${{s}}$ and outputs ${\widehat{s}}$. 
Then, with the latent vector for each input sequence, the similarity of each pair of sequences can be calculated with a certain distance metric (e.g., Euclidean or cosine distance).

However, both types of methods above only consider the order of events within a sequence. 
For example, these methods do not consider the time interval between events. 
Medical records of different patients often have different time intervals. 
Also, records of one patient often have varied time intervals. 
Another example is that these methods do not account for the correlations between events within the same patient's records. 
Two different events that happened at different times may have a high correlation (e.g., a patient may have recurring symptoms after a certain period of time).
Based on these observations, we extend the LSTM autoencoder framework by integrating the self-attention mechanism~\cite{lin2017structured} to better capture the medical records' temporal patterns.
\section{Background and Analysis Targets}
\label{sec:background}

This section describes the characteristics of EMR data and analytic targets,  
which dictate the requirements for our model and visualization system.

\subsection{General Characteristics of EMR Data}
\label{sec:data_characteristics}

As described in~\cite{lee2017big}, the general characteristics of EMR data make data analysis challenging. 
We describe three critical aspects we address in this work.

\vspace{5pt}
\begin{compactdesc}
    \item[High-dimensionality]
    EMR data typically consists of a large number of medical features, such as multiple medical tests (e.g., serum calcium concentration), medications, diagnoses, and procedures. 
    \item[Irregularity in time]
    The irregularity of EMR data is caused by the fact that each patient's medical features are recorded only when they visit the hospital. 
    As a result, each patient's records, which can be represented as a temporal sequence, have different intervals between each pair of events and also often have different lengths.
    \item[High portion of missing data \& data sparsity]
    EMR data often suffers from a high proportion of missing data.
    This can be caused by either data collection problems (i.e., patients are only checked for certain medical considerations) or documentation problems (i.e., machine breakdowns or human-entering errors~\cite{wells2013strategies}). 
    Apart from a high proportion of missing data, data sparsity is another general characteristic of EMR. 
    Sparsity is unavoidable since most patients visit the hospital only a few times and  they usually take only a small subset of medical examinations and treatments.
\end{compactdesc}

\subsubsection{Description of Data Used in Our Study}
\label{sec:our_data}
The dataset used in our study contains medical test records for 854 neonates in the Neonatal Intensive Care Unit (NICU) from an academic regional medical center in the United States. 
Their records were collected during the years 2015 and 2016.
Each neonate's records consist of laboratory testing results performed during their multiple hospital visits and NICU stays. 
While 239 different medical tests are recorded, each neonate took a subset of these medical tests per hospital visit (45 medical tests on average).
Therefore, this data also shares the same general characteristics of EMR data (i.e., 239 dimensions, different time spans and lengths, and high sparsity). 
We use this dataset as an example of EMR data through our work. 
However, we should note that our model and visual analytics system are designed to be applicable to other EMR datasets as well.

\subsection{Target Analysis Tasks}
\label{sec:targets}

Our work's general target is to support the analysis of similar patients' medical records. 
As mentioned, analyzing similar records is particularly important when the clinicians make decisions for their patients~\cite{jin2019carepre}.
Under this general target, we set several detailed analysis targets below. 

\textbf{T1: Find similar medical records.} The most fundamental task we need to support is finding similar medical records. 
Based on the user-selected focal patient's records, the system should provide the most similar ones (e.g., top-3 similar records) from other patients' records.

\textbf{T2: Find (dis)similar time points among similar records.} 
Even though the overall similarities are high between the focal patient's and similar patients' records, the similarities might vary along with their clinical progressions. 
Especially, finding dissimilar time points is important for medical decisions.
For example, the focal patient may have had a sudden increase in blood pressure and could not be treated with specific medicines used to treat the other similar patients. 

\textbf{T3: Compare medical feature values at a certain time point}. 
This task is to supplement T2. 
As mentioned in the example in T2, the user would want to know the reason why the patients are similar or different at some specific time points. 

\textbf{T4: Understand general tendencies in similar records' medical feature values.}
Another important task is, between the similar records, which and how medical features are generally (dis)similar across time points.
For example, through this task, the user could find all of these patients relatively high lymphocytes while only the focal patient has low red cell counts in his/her blood. 

\textbf{T5: Compare specific medical feature's values across time.} 
Once the user finds a medical feature of which the focal patient has a different tendency (e.g., low red cell counts) through T3 or T4, the user would want to know more details about the corresponding feature (e.g., whether the patient's red cell counts are constantly low or rapidly decreased at some time point). 

First, in \autoref{sec:model}, we introduce a method for calculating the similarity of medical records, which is necessary to support T1. 
Then, in \autoref{sec:vis}, we describe our visual analytics system supporting all the tasks above.

\section{Similarity Calculation with Sequence Embedding}
\label{sec:model}

\begin{figure}[t]
    \centering
    \includegraphics[width=1.0\linewidth]{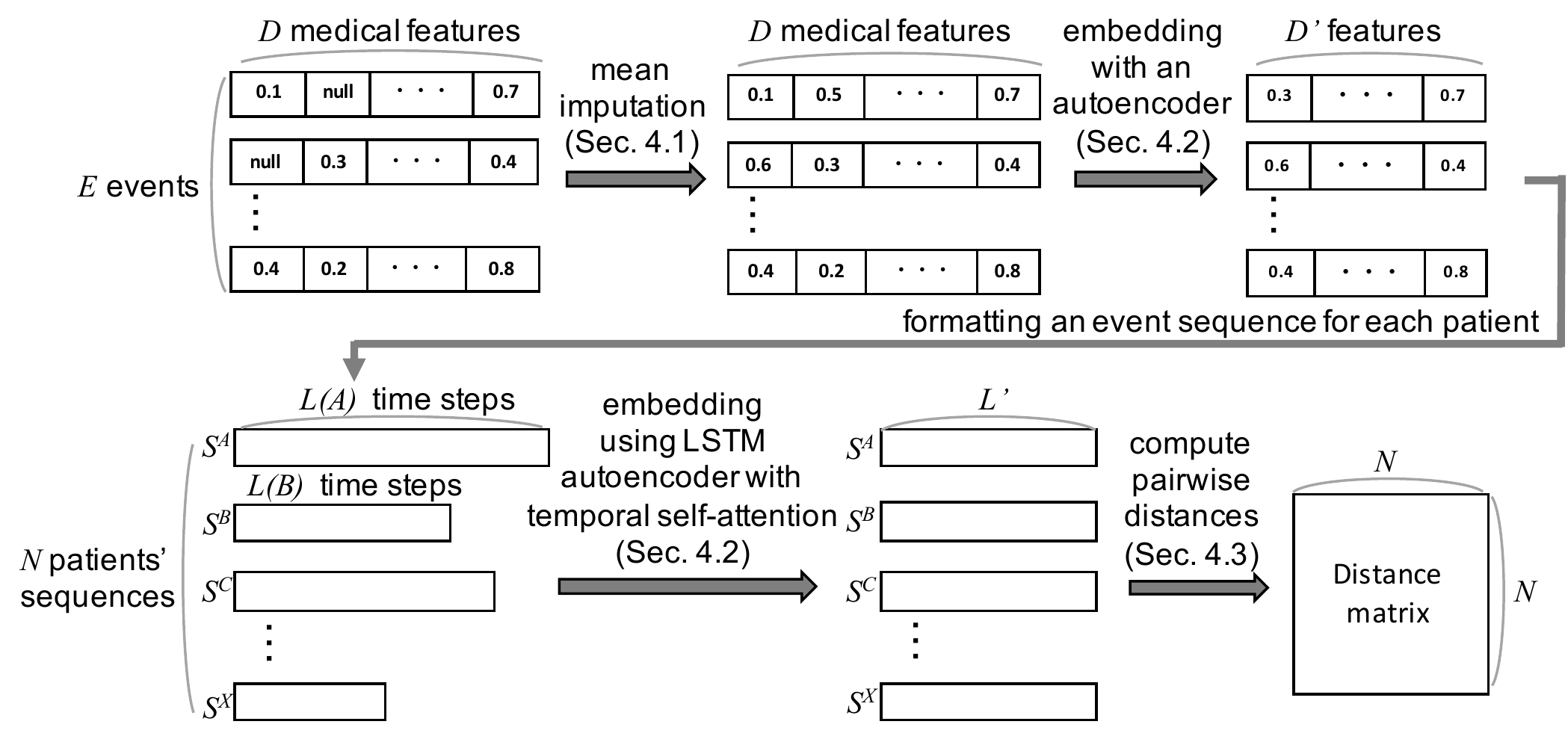}
    \caption{The overview of similarity calculation steps.}
    \label{fig:embedding_flow}
\end{figure}

We present our methodology for similarity calculation of medical records with different lengths in this section. 
An overview of the flow of our similarity calculation is shown in \autoref{fig:embedding_flow}.
We use two-step embeddings to convert medical records to fixed-length vectors. 
The first step is (1) \textit{event embedding}, where each event consisting of multiple medical feature values is converted into a vector.
Then, we apply (2) \textit{sequence embedding}, where each sequence consisting of multiple embedded vectors obtained in (1) is converted into a vector.
In order to deal with the high-dimensionality and sparsity of the data (refer to \autoref{sec:background}), we use the first step to learn lower-dimensional latent representations of the original features. 
The second step is used to handle the irregularity of the data.
Lastly, after obtaining fixed-length vectors with these two steps, we compute the similarity of each pair of these vectors with a certain distance metric, such as the Euclidean distance.

While the existing work for analyzing medical records~\cite{guo2018visual,jin2019carepre} also embedded each event to a latent vector, there are two major differences with our model. 
First, they handled only medical events (e.g., diagnosing insulin) without additional information (e.g., how much insulin is diagnosed). 
This information is important to analyze many of the medical records.
For example, in our dataset, while the neonates took a subset of tests at each recorded time, we should consider not only the information of which tests they took (e.g., measuring the amount of lymphocytes in blood) but also the information of the tests results (e.g., 1,000 lymphocytes per $\SI{1}{\micro\liter}$). 

Furthermore, Jin et al.~\cite{jin2019carepre} calculated the similarity of two medical records with DTW after the event embedding instead of going through our second embedding step.
As discussed in \autoref{sec:relatedmodel}, DP-based distance measures, including DTW, has a limited capacity to deal with the irregularity of medical records. 

\subsection{Data Prepossessing}
\label{sec:data_preprocessing}
EMR data often has a high proportion of missing values and/or high sparsity. 
Therefore, before applying embeddings, we need to fulfill the empty values. 
We adopt mean imputation~\cite{rubin1987multiple}, which is the most widely-used method. 
By learning low-dimensional latent representations from the imputed results with the embedding described in \autoref{sec:event_embedding}, the influence from the mean imputation on the calculation of the similarity of medical records can be moderated. 

However, we should note that mean imputation still may bring a problem of underestimated variability. 
In our case, this problem causes the possibility that our method judges two neonates who have taken a quite different set of tests to have high similarity. 
Rather than designing a complex algorithm for this problem, we provide an interface for interactive filtering based on the similarity of tests taken, as described in \autoref{sec:simtest}.

\subsection{Event Embedding}
\label{sec:event_embedding}

For the first embedding, we use a basic autoencoder~\cite{kramer1991nonlinear}, a neural network consisting of three layers (i.e., input, hidden, and output layers). 
We use this embedding to compress each event's high-dimensional medical features to a lower-dimensional representation.
Let $D$ be the number of medical features stored in medical records. 
The event embedding produces $D' (D' \ll D)$ medical features for each event (i.e., each time step). 
This is essentially a similar approach to applying principal component analysis (PCA)~\cite{jolliffe1986principal} for reducing the total number of features to avoid the `curse of dimensionality'~\cite{verleysen2005curse}.
As the original purpose of the autoencoder~\cite{kramer1991nonlinear}, it performs nonlinear PCA on a dataset. 
This is more suitable for EMR data due to its high dimensionality and complexity.

In EMR data, for example, each event corresponds to a unique hospital visit by one patient and consists of multiple medical test results. 
Let ${e}_{i}^{A}$ be the $i$-th event of patient $A$, ${u}_{i}^{A}$ be the latent vector representation of ${e}_{i}^{A}$ after the embedding (the length of ${u}_{i}^{A}$ is $D'$), and $\tau_{i}^A$ be the time when event ${e}_{i}^{A}$ occurred.
Now, the original sequence of patient $A$ is converted to sequence ${s}^A$ consisting of multiple event vectors 
\begin{equation}
  {s}^A = \langle({u}_{1}^A, \tau_{1}^A),({u}_{2}^A, \tau_{2}^A), \cdots, ({u}_{L(A)}^A, \tau_{L(A)}^A)\rangle,
  \label{eq:sequence}
\end{equation}
where $L(A)$ is the length of sequence ${s}^A$.

For our dataset of the neonates' medical test records, in total, 239 different types of medical tests are recorded (i.e., $D=239$). 
Then, by embedding these events, we obtained latent vectors of size 32 (i.e., $D'=32$) for each time point. 

\subsection{Sequence Embedding}
\label{sec:seq_embedding}
 
For the second embedding, we use the seq2seq framework~\cite{sutskever2014sequence}. 
As explained in \autoref{sec:relatedmodel}, the seq2seq framework has been developed to generate an output sequence from an input sequence. 
For example, an English sentence and a translated French sentence could be input and output sequences, respectively. 
One major strength of the seq2seq framework is that the input (and/or output) sequences could have any different lengths because it uses RNNs as both encoder and decoder. 
Specifically, we use RNNs with LSTM units to handle long-range temporal dependencies. 
We call the seq2seq with LSTM units as the LSTM autoencoder in the ensuing description.
In our case, with the LSTM autoencoder, we obtain the fixed-length latent vector converted from the different length inputs (i.e., medical records).

Additionally, we incorporate the self-attention mechanism~\cite{bahdanau2015neural,lin2017structured} into the LSTM autoencoder. 
The self-attention mechanism is known for improving the learning performance of seq2seq~\cite{bahdanau2015neural}.
In addition to this reason, more importantly, we use the self-attention mechanism to model the time interval between events and correlations among events. 
To achieve this, we enhance the self-attention mechanism by providing control over the attention weight based on the time interval. 
We name this enhanced self-attention mechanism as the \textit{temporal} self-attention.

\begin{figure}[tb]
    \captionsetup{farskip=0pt}
	\centering
	\subfloat[Original model in \cite{bahdanau2015neural}]{
     \includegraphics[width=0.45\linewidth]{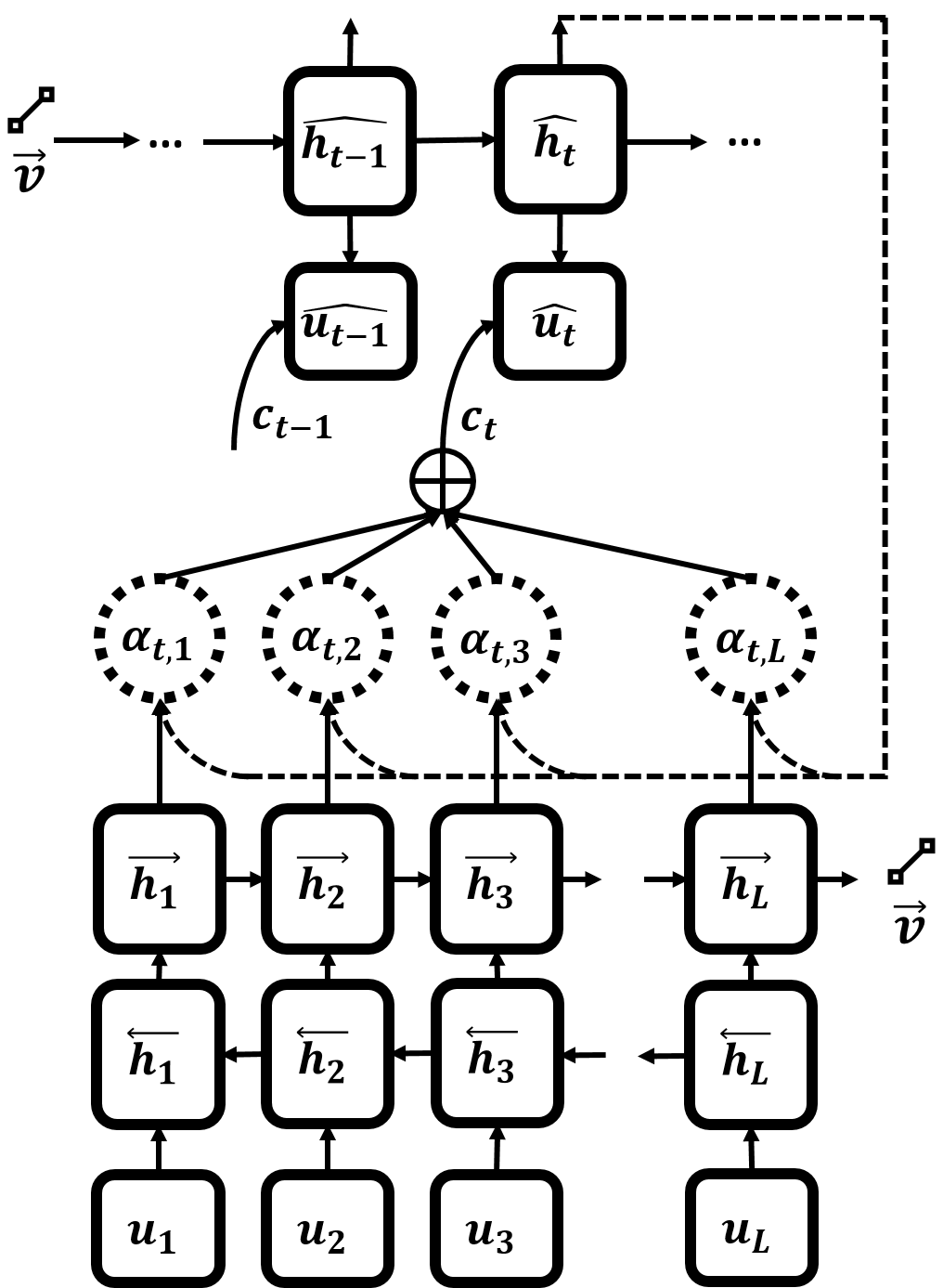}
     \label{fig:attention_a}
    }
    \subfloat[Our model]{
     \includegraphics[width=0.45\linewidth]{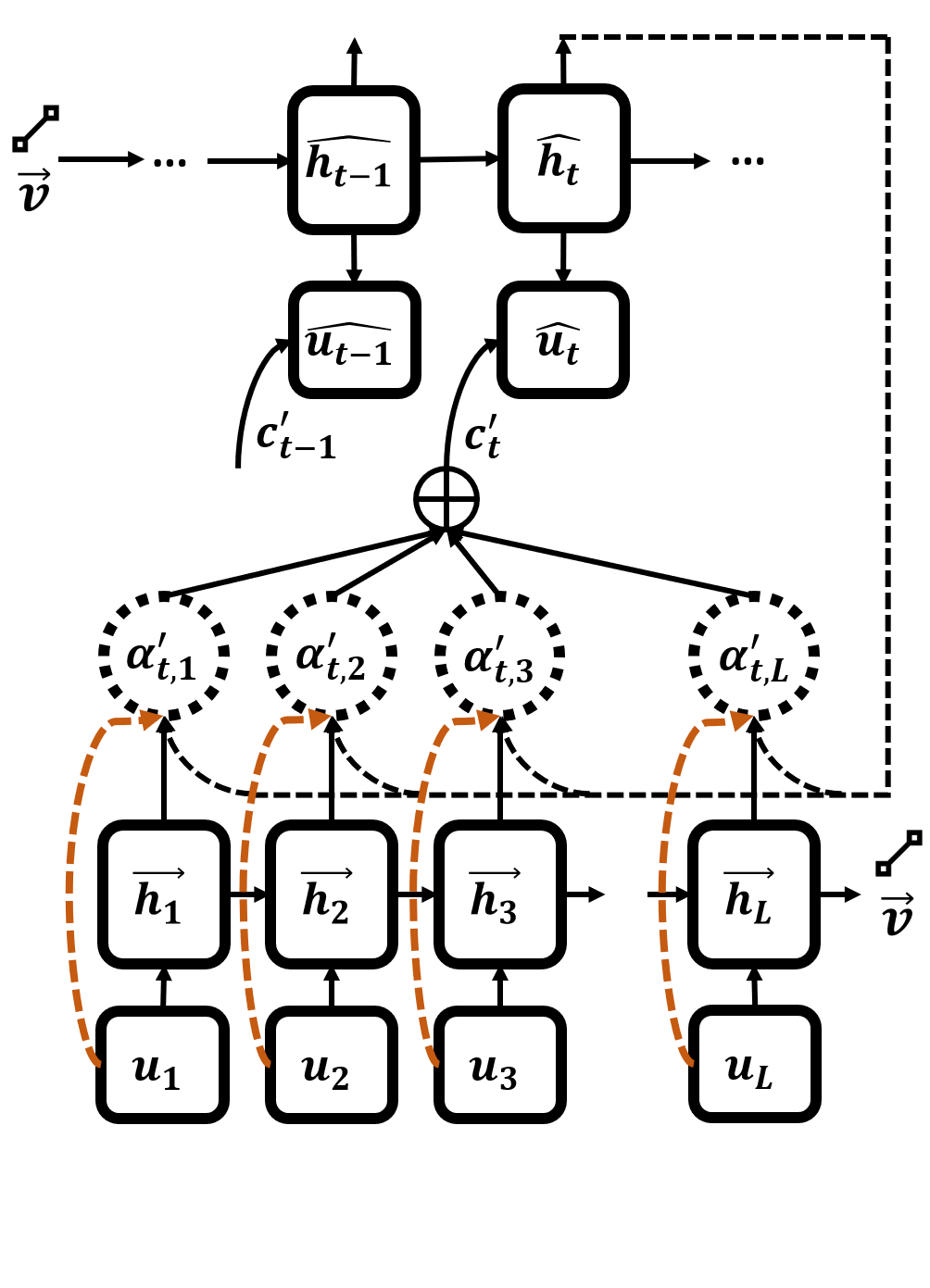}
     \label{fig:attention_b}
    }
   	\caption{Two different models of the LSTM autoencoder with the self-attention mechanism. Orange arrows show the inputs of time interval information that we incorporate in the temporal self-attention.}
	\label{fig:attention}
\end{figure}

\begin{figure*}[tb]
    \captionsetup{farskip=0pt}
	\centering
	\subfloat[The results for ID 80]{
     \includegraphics[width=0.8\linewidth]{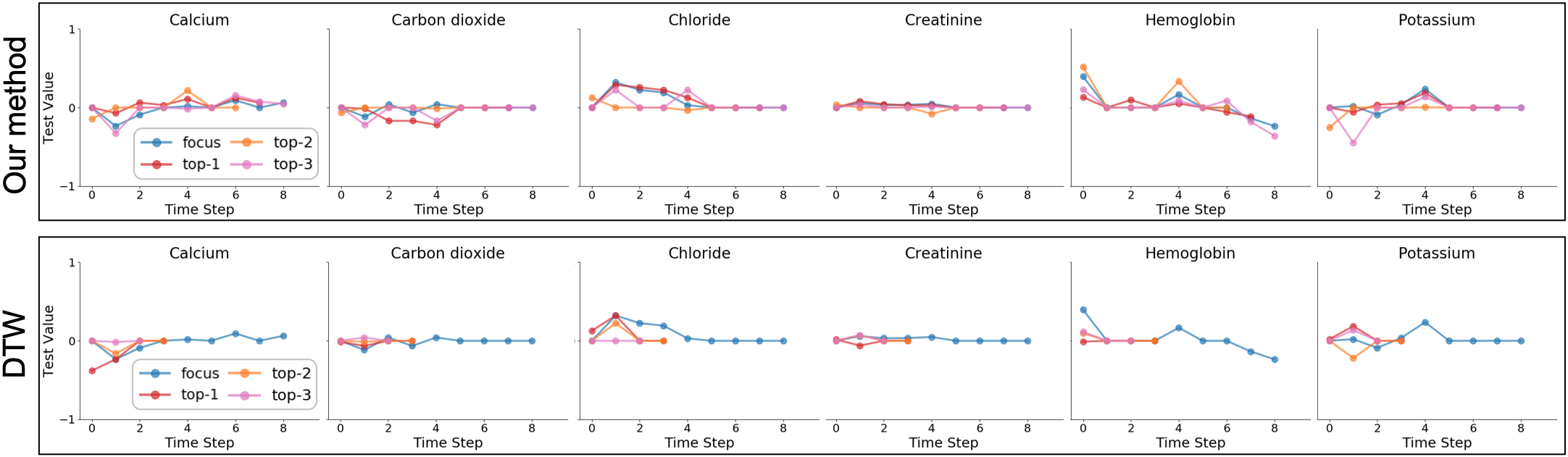}
     \label{fig:query1}
    }\\
    \vspace{8pt}
    \subfloat[The results for ID 21]{
     \includegraphics[width=0.8\linewidth]{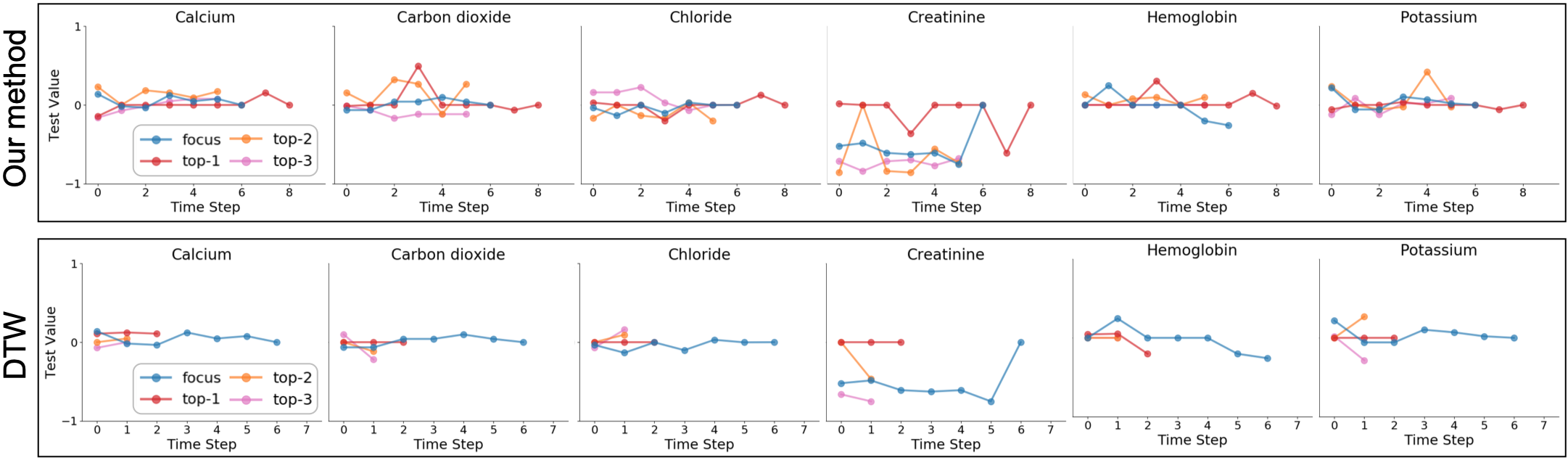}
     \label{fig:query2}
    }
  	\caption{The comparison of the top-3 similar neonates' records identified with our method (top row) and DTW (bottom row). We select (a) ID 80 and (b) ID 21 as two examples of focal neonates. Each column corresponds to each of 6 different test items selected from 239 randomly.}
	\label{fig:performance}
\end{figure*}

\subsubsection{LSTM Autoencoder with the Self-Attention Mechanism}
Before our temporal self-attention, we first provide a brief introduction of the original self-attention mechanism used in the LSTM autoencoder~\cite{bahdanau2015neural} shown in \autoref{fig:attention_a}.
In \autoref{fig:attention_a}, ${U}=\{u_1, u_2, \cdots, u_L\}$, $\hat{U}=\{\hat{u}_1, \hat{u}_2, \cdots, \hat{u}_L\}$, and $\Vec{v}$ represent input sequences, output sequences, and embedded sequences respectively.
Specifically, the input sequence $U$ is the medical time sequences after the first embedding $\langle u_1, u_2, \cdots, u_L\rangle$ without time information $\langle \tau_{1}, \tau_{2}, \cdots, \tau_{L}\rangle$; the output sequence $\hat{U}$ is the reconstruction of $U$; and the embedded sequence $\Vec{v}$ is the fixed-length vector representation of the input sequence that we want to obtain.

To predict the $t$-th vector of output $\hat{u}_t$, the model refers to $\hat{u}_{t-1}$, the LSTM autoencoder's $t$-th hidden state $\hat{h}_t$, and the fixed-length latent vector $c_t$, called the context vector. 
With the self-attention mechanism, $c_t$ is computed with a sequence of ``annotations''~\cite{bahdanau2015neural} $\{h_1, \cdots, h_L\}$.
Each annotation $h_i$ is a hidden state computed from an input sequence $U$ and contains information about $U$ with a strong focus on the parts surrounding 
$u_i$.
$c_t$ can be computed as a weighted sum of a sequence of annotations $h_i$:
\begin{equation}
    c_t = \sum_{i=1}^{L} \alpha_{t,i} h_i
\end{equation}
where $\alpha_{t,i}$ is the weight of annotation $h_i$.
Specifically, $\alpha_{t,i}$ is computed by
\begin{equation}
    \alpha_{t,i} = \frac{\exp{(b_{t,i})}}{\sum_{k=1}^{L}\exp{(b_{t,k})}}
    \label{eq:original_alpha}
\end{equation}
where
\begin{equation}
    b_{t,i} = f(\hat{h}_{t-1}, h_i).    
    \label{eq:original_alpha_2}
\end{equation}
Here, $f(\cdot)$ is a neural network generating $b_{t,i}$ from hidden state $\hat{h}_{t-1}$ and annotation $h_i$.
Essentially, $\alpha_{t,i}$ shows how much attention the model should pay to the information around $u_i$ when it predicts the $t$-th output vector $\hat{u}_t$.
Refer to \cite{bahdanau2015neural} for more details about their model.

\subsubsection{Temporal Self-Attention}
\autoref{fig:attention_b} shows the LSTM autoencoder with the temporal self-attention. 
A fundamental difference from the model in \cite{bahdanau2015neural} is using $\alpha'_{t,i}$ instead of using $\alpha_{t,i}$ in \autoref{eq:original_alpha}.
Our model computes the weight $\alpha'_{t,i}$ with consideration of the time interval between the $i$-th input and $t$-th output events. 
That is, 
\begin{equation}
    \alpha'_{t,i} = \frac{\exp{(b'_{t,i})}}{\sum_{k=1}^{L}\exp{(b'_{t,k})}}
    \label{eq:our_alpha}
\end{equation}
where
\begin{equation}
    b'_{t,i} = f(\hat{h}_{t-1}, h_i, d_{t,i}).
    \label{eq:our_alpha_2}
\end{equation}
As seen in the difference between \autoref{eq:original_alpha_2} and \autoref{eq:our_alpha_2}, our model uses the time interval $d_{t,i}$ between the $i$- and $t$-th events, when calculating the weight $\alpha'_{t,i}$.
We compute $d_{t,i}$ with $\tau_{t}$ and $\tau_{i}$ in \autoref{eq:sequence}:
\begin{equation}
    d_{t,i} = \left\lfloor \left| \tau_{t} - \tau_{i} \right| \right\rfloor,
    \label{eq:time_interval}
\end{equation}
where $\lfloor \cdot \rfloor$ is a floor function to convert the real value to an integer. To control the precision of the difference of time intervals, the unit of $d_{t,i}$ (e.g.., a second, hour, or day) should be decided, depending on the type of medical records or analysis target.
Now, with \autoref{eq:our_alpha}, \ref{eq:our_alpha_2}, and \ref{eq:time_interval}, the model can consider the time interval and/or the temporal correlation among events during the sequence embedding.

Lastly, we use the output of the last encoder's hidden state as the embedded fixed-length vector $\Vec{v}$ of the original input sequence, which is also an input vector to the first decoder cell. The training objective function of the model is to minimize the error between the input and output sequences as $\min \|\hat{U} - {U}\|_2$.

Our model has another difference with the original model in~\cite{bahdanau2015neural}.
While the model in \cite{bahdanau2015neural} used a bidirectional RNN~\cite{schuster1997bidirectional}, we use a one-directional LSTM to keep the model simple and to compute the result faster. 
However, this is a minor difference and we can replace it with a bidirectional RNN if necessary.

For our dataset, with the sequence embedding step in this section, we have converted each of the 854 neonates' records into a fixed-length vector with a size of 128. 

\subsection{Similarity Calculation with Embedded Vectors}

From the above two embedding steps, we have a fixed-length embedded vector for each patient's records. 
Now, we can compute the sequence similarities based on these vectors with a certain distance metric, such as the Euclidean distance or cosine distance.
For out dataset, we tested both the Euclidean and cosine distances and the resultant performance was similar; thus, we simply selected the Euclidean distance in our system.

\subsection{Comparison with Dynamic Time Warping}

To demonstrate the effectiveness of our method for calculating the similarity of medical records with different lengths, we compare our method with a baseline similarity measurement: the Dynamic Time Warping (DTW)~\cite{DTW}. 
We perform the task of identifying similar records in the NICU dataset described in \autoref{sec:our_data}. 
Specifically, we use both methods to find the top-3 similar sequences to the given focal neonate. 
Then, we visually compare the focal and identified similar records to see which method provides more reasonable results.

\autoref{fig:performance} shows the results for two different focal neonates (ID 80 and ID 21)\footnote{We only present the results for two different focal neonates here. The same task is performed on all the neonates' records except for the records that have less than two events. All results similar to \autoref{fig:performance} are available at \url{https://drive.google.com/drive/folders/1ypDt00VJKZqvSgW8GFcHtZUjDfZcAagM?usp=sharing}.}.
Because the NICU dataset is multivariate data consisting of 239 different medical tests, instead of displaying temporal sequences for each of all medical tests, we randomly choose 6 tests (Calcium, Carbon dioxide, Chloride, Creatinine, Hemoglobin, and Potassium) to make it possible to visually compare and describe the concrete differences from the results.
In \autoref{fig:performance}, we have not applied any sequence alignment to avoid including the effect from the alignment. 
In \autoref{fig:performance}a and b, the first and second rows show the results using our model and DTW, respectively.

From these results, we can see that our method generally selects temporal sequences that take similar values and patterns with the focal neonates. 
For example, as for Calcium in \autoref{fig:query1}, all sequences have ``valleys'' (i.e. lower values than other close time steps) around the beginning and ``peaks'' (i.e., higher values than other close time steps) around the middle time steps. 
Another clear example can be seen in Hemoglobin of \autoref{fig:query1}. 
The values in all sequences tend to decrease as time progresses but have some peaks around the middle. 
On the other hand, we can see that the similar sequences selected by DTW tend to be short and do not show the same patterns with the focal neonate.
For example, in Carbon dioxide of \autoref{fig:query2}, while the focal sequence has a slight increase in its values, DTW selects two short decreasing lines as the top-2 and -3 similar sequences. 
This could happen because, during the similarity calculation, DTW tries to find the best alignment between two sequences based on simple dynamic programming. 
Then, if many time steps have less dissimilarity to the corresponding aligned time step, the total dissimilarity would be small. 

In summary, the results above show the strength of our method when compared with DTW.
The strength is derived from using deep neural networks to capture the overall transition and temporal patterns of sequences, instead of merely focusing on comparing individual event pairs.

\begin{figure*}[t]
    \centering
    \includegraphics[width=1.0\linewidth]{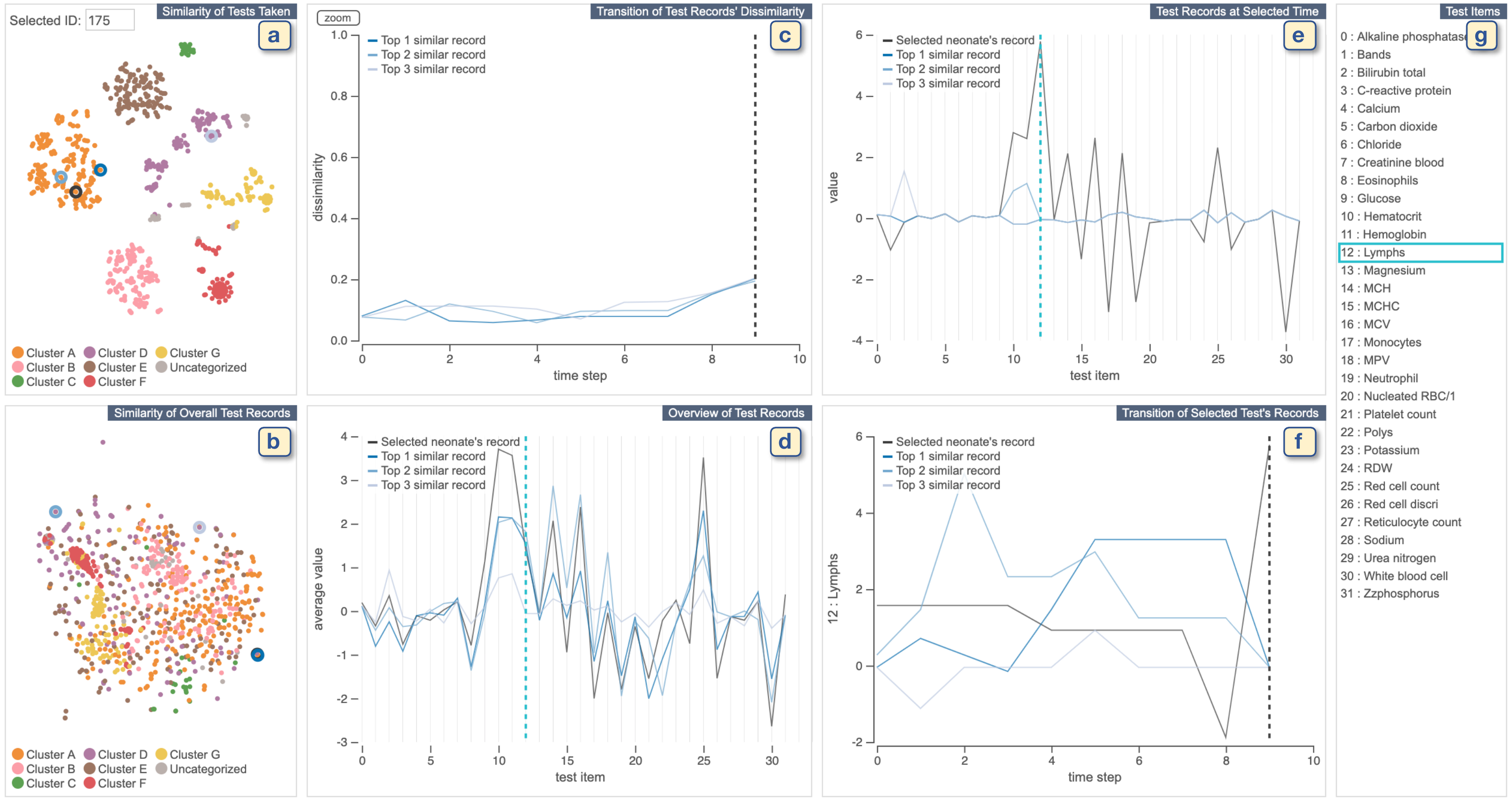}
    \caption{The user interface of our visual analytics system. The system consists of multiple views. 
    (a) and (b) show each neonate's similarities of medical tests taken and the records of test values, respectively.
    (c) depicts changes of test records dissimilarities over time between the neonate selected from (a) or (b) and top-3 similar neonates. 
    (d) visualizes a statistical overview of the selected and top3-similar neonates. 
    (e) provides all the medical test results at the selected time in (c) or (f). 
    (f) shows the temporal changes of values of the selected test in (d) or (e). 
    Finally, (g) lists all medical test names.}
    \label{fig:system_overview}
\end{figure*}

\section{Visual Analytics System}
\label{sec:vis}

We present a visual analytics system that is designed to support the analysis tasks described in \autoref{sec:targets}. 
We first describe an overview of the system with an analysis workflow and then details of each view of the system. 
We use our dataset of the neonates to make the system description more concrete.
However, the system can be applied to other datasets consisting of multivariate values for each event. 

\begin{figure}[ht]
    \centering
    \includegraphics[width=1.0\linewidth]{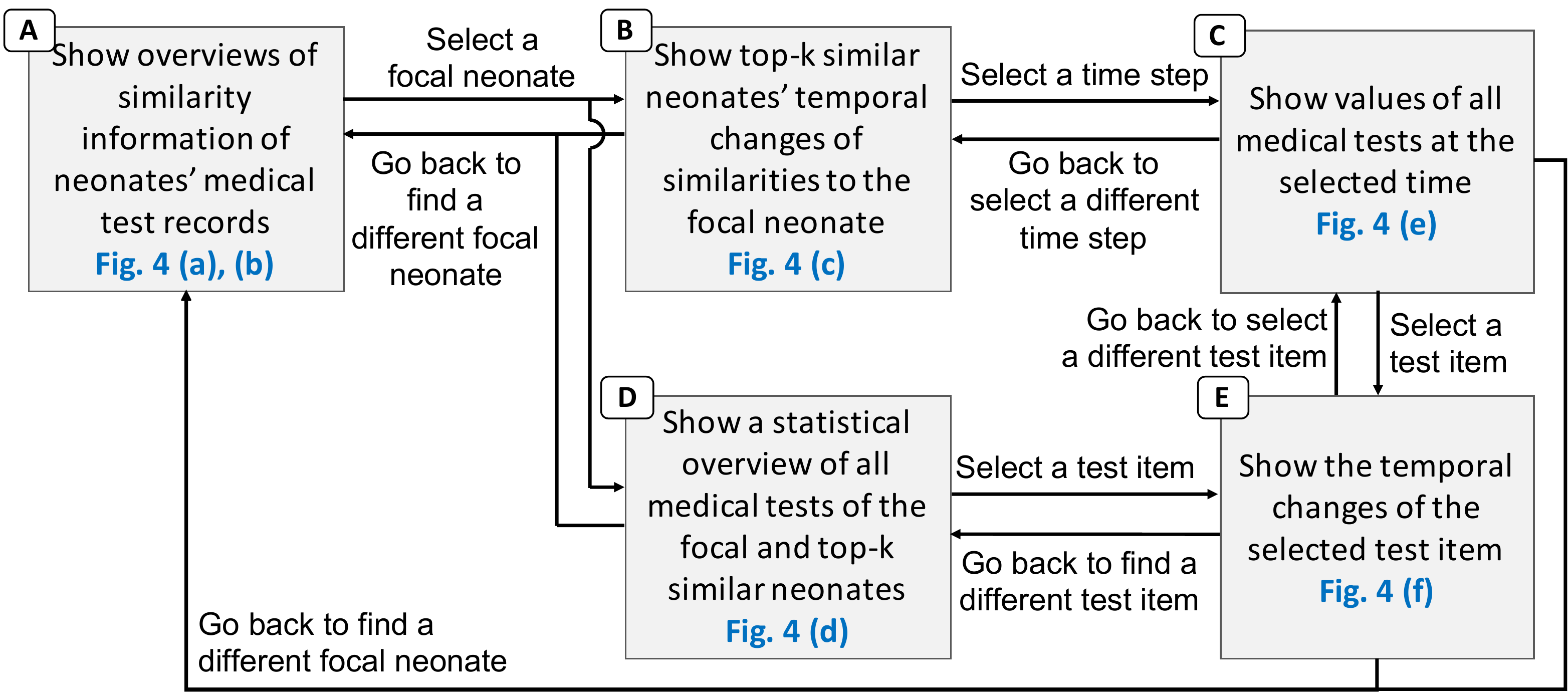}
    \caption{The analysis workflow of using our visual analytics system.
    }
    \label{fig:analysis_workflow}
\end{figure}

\subsection{System Overview and Analysis Workflow}
We describe an overview of our visual analytics system shown in 
\autoref{fig:system_overview} by going through the explanation of the analysis workflow in \autoref{fig:analysis_workflow}.
First, the user (e.g., a clinician) can see the overviews of similarity information of neonates' medical test results (\autoref{fig:analysis_workflow}A). 
Our system provides two overviews of all neonates: one is for neonates' similarities based on the combination of tests taken during the recorded period (\autoref{fig:system_overview}a) and another is neonates' similarities based on the test values (\autoref{fig:system_overview}b). 
Then, by referring to these views, the user selects one focal neonate. 
For example, in \autoref{fig:system_overview}a, from the orange cluster, the neonate highlighted with the black outer-ring is selected.

Based on the focal neonate, the system automatically selects the top-$k$ similar neonates (specifically, $k=3$ as a default) and then visualizes overviews of the focal and top-$k$ neonates, as shown in \autoref{fig:system_overview}c and \autoref{fig:system_overview}d (corresponding to \autoref{fig:analysis_workflow}B and D). 
Because medical records can be represented as temporal multivariate event sequences, to help the user decide a time step or a variable to be reviewed, in \autoref{fig:system_overview}c and \autoref{fig:system_overview}d, we provide overviews from temporal and variable aspects, respectively. 
More specifically, while \autoref{fig:system_overview}c shows temporal changes of dissimilarities of the top-$k$ neonates' test results to the focal neonate, \autoref{fig:system_overview}d visualizes the statistical summary of each test's values across the recorded time period. 

When the user wants to review the test values in detail based on the temporal changes, the user can select one time step from the view at \autoref{fig:system_overview}c. 
For example, in \autoref{fig:system_overview}c, the last time step is selected because the dissimilarity goes higher. 
Then, to review why they are (dis)similar, the system shows each neonate's all test values at the selected time, as shown in \autoref{fig:system_overview}e (corresponding to \autoref{fig:analysis_workflow}C). 
Furthermore, from the result shown in \autoref{fig:system_overview}e, the user can select a specific test item in which he/she wants to see temporal changes in detail. 
For example, in \autoref{fig:system_overview}e, we select the test item `12: Lymphs' because its values have the largest difference between the focal and other neonates. 
The result is visualized in \autoref{fig:system_overview}f (corresponding to \autoref{fig:analysis_workflow}E). 
From \autoref{fig:system_overview}f, for example, the user can understand the focal neonate's high `lymphs' at the last time step have increased the dissimilarity observed in \autoref{fig:system_overview}c.

On the other hand, when the user starts to review the details from the overview of variables visualized in \autoref{fig:system_overview}d (corresponding to \autoref{fig:analysis_workflow}D), the user can select a test item of interest from the result in \autoref{fig:system_overview}d and the details will be displayed in \autoref{fig:system_overview}f. 

While we have described the two major analysis flows above,  \autoref{fig:analysis_workflow}A $\rightarrow$ B $\rightarrow$ C $\rightarrow$ E (focusing on temporal changes) and A $\rightarrow$ D $\rightarrow$ E (focusing on differences across test items), each of these analysis steps is tightly connected and, in practice, we often go back and forth across these different steps. 
Also, we want to note that the arrangement of the views in our system is designed to match the order of the two major analysis flows. 

\subsection{Similarity of Tests Taken (\autoref{fig:system_overview}a)}
\label{sec:simtest}

The views in \autoref{fig:system_overview}a and \autoref{fig:system_overview}b are used to perform T1 in \autoref{sec:targets} with our similarity calculation described in \autoref{sec:model}.
As discussed, EMR data, including our neonate dataset, is often sparse. 
We have addressed this problem to some extent with the similarity calculation using the two-step embedding.
However, due to the mean imputation in the data preprocessing step, as described in \autoref{sec:data_preprocessing}, our method still may judge two neonates who have taken a quite different set of tests to have high similarity. 
\autoref{fig:system_overview}a can be used to deal with this problem.

In this view, we visualize similarities of the neonates' records based on test items they took during the collected time period. 
First, we employ the Jaccard index~\cite{jaccard} to compute the similarity.   
Then, to visualize all the neonates' similarity relationships in a single 2D plot, we apply a dimensionality reduction (DR) method, specifically t-SNE~\cite{maaten2008visualizing}. 
Afterwards, to extract clusters from the DR result, we use HDBSCAN~\cite{Campello2013HDBSCAN}, one of the density-based clustering methods.
Lastly, to inform the clustering information, we color each point (i.e., each neonate's record) based on the assigned cluster-ID. 
We use categorical colors with enough differences in their hues to distinguish each cluster. 
Also, because the density-based clustering, including HDBSCAN, would not assign some points (e.g., outliers or noises) to any specific cluster, we color such points with gray and label them as `uncategorized'.

The user can select a focal neonate from the input dialog placed on the top left (e.g., ID 175 is selected in \autoref{fig:system_overview}a) or by clicking a point in the view. 
We indicate the focal neonate with a black outer-ring, as shown in \autoref{fig:system_overview}a.
Then, the system automatically searches for the top-3 similar neonates' records based on the pre-computed similarities with the method described in \autoref{sec:model} (note that these similarities are based on the test values).
The selected top-3 neonates are indicated with outer-rings of blue colors with different saturation.
The darker blue represents a neonate with a higher similarity. 
We have chosen this single hue encoding to ensure these colors do not share the same hue with the cluster colors to avoid misleading the user.
From an example of \autoref{fig:system_overview}a, we can see that while the focal and first- and second-top neonates are selected from the orange cluster (i.e., Cluster A), the third-top neonate is selected from the purple cluster (i.e., Cluster D). 
This is an example of the problem stemming from mean imputation.
In this case, the user might want to select similar neonates only within the orange cluster.

\begin{figure}[tb]
    \captionsetup{farskip=0pt}
	\centering
	\subfloat[Before selecting Cluster A]{
     \includegraphics[width=0.4\linewidth]{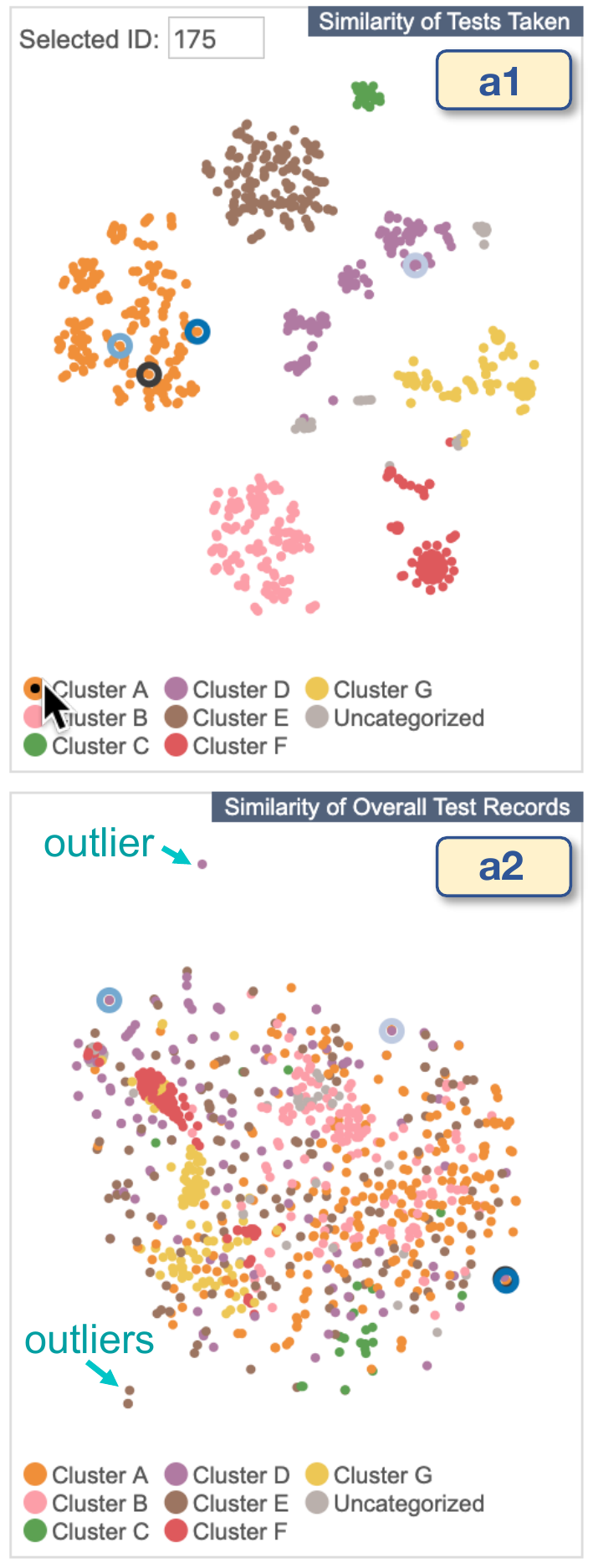}
     \label{fig:cluster_filtering_a}
    }
    \subfloat[After selecting Cluster A]{
     \includegraphics[width=0.4\linewidth]{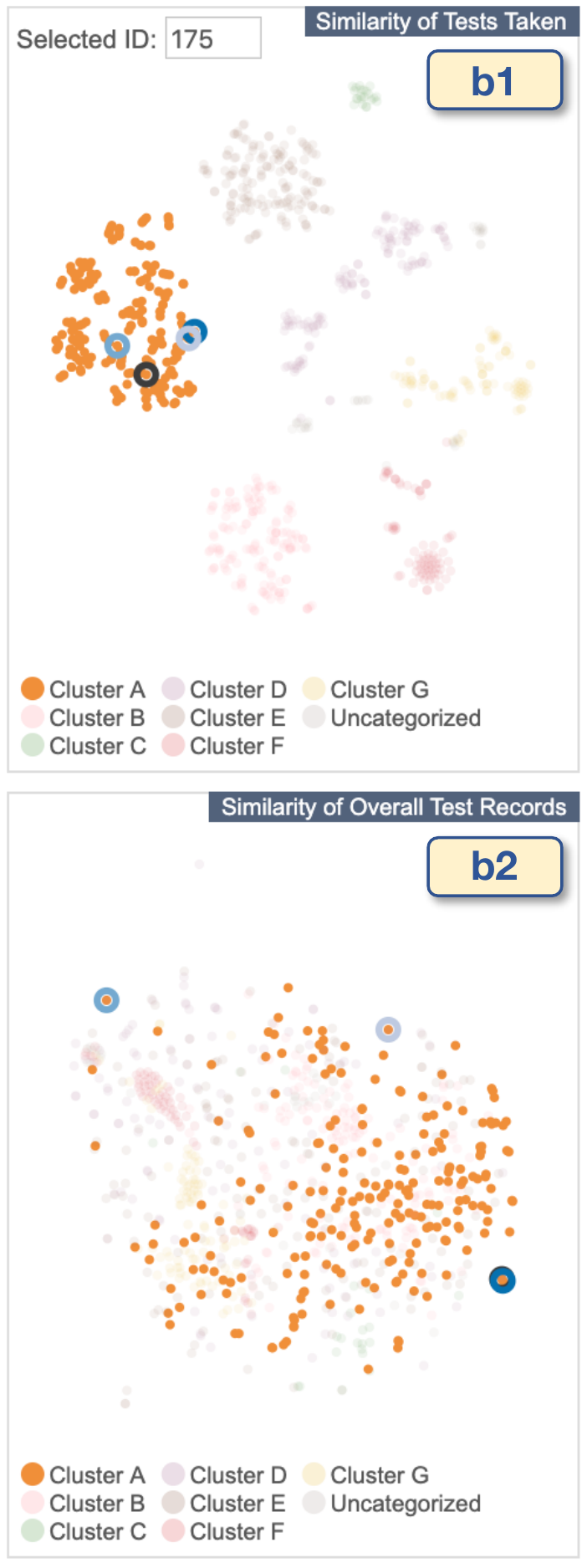}
     \label{fig:cluster_filtering_b}
    }
   	\caption{The selection of similar neonates within an indicated cluster.}
	\label{fig:cluster_filtering}
\end{figure}

To support such selection, we provide an interaction to restrict the identification of similar neonates only within the selected cluster. 
An example of the interaction result is shown in \autoref{fig:cluster_filtering}. 
First, the user can specify a certain cluster from the legend of clusters placed at the bottom of the view.
In \autoref{fig:cluster_filtering}a1, the mouse-hovered cluster is indicated with a dot at the center of the circle (analogous to the design of the commonly-used radio button).
Then, the user can select a cluster by mouse-clicking. 
The system immediately selects and updates the similar neonates only within the selected cluster if the focal patient belongs to the selected cluster, as shown in \autoref{fig:cluster_filtering}b1.
Additionally, the system highlights the selected cluster by reducing the opacity for others. 
Also, we allow the user to update a focal neonate after the cluster selection. 
In this case, the system selects similar neonates within the selected cluster. 
The user can cancel the selection of the cluster by clicking on the selected cluster again. 

\subsection{Similarity of Overall Test Records (\autoref{fig:system_overview}b)}
\label{sec:similarity_test_records}

This view shows an overview of the similarities of all the neonate test records. 
We compute the similarity of each pair of the test records with the method described in \autoref{sec:model}.
Then, similar to \autoref{fig:system_overview}a, we apply t-SNE to visualize these similarity relationships. 
By providing this overview, the user can find specific groups in which all the neonates have similar test values (i.e., identifying cohorts of neonates) or outliers from other neonates.
For example, in \autoref{fig:cluster_filtering}a2, we can see three neonates are placed far from the others, as indicated with teal arrows.

\begin{figure}[t]
    \centering
    \includegraphics[width=\linewidth]{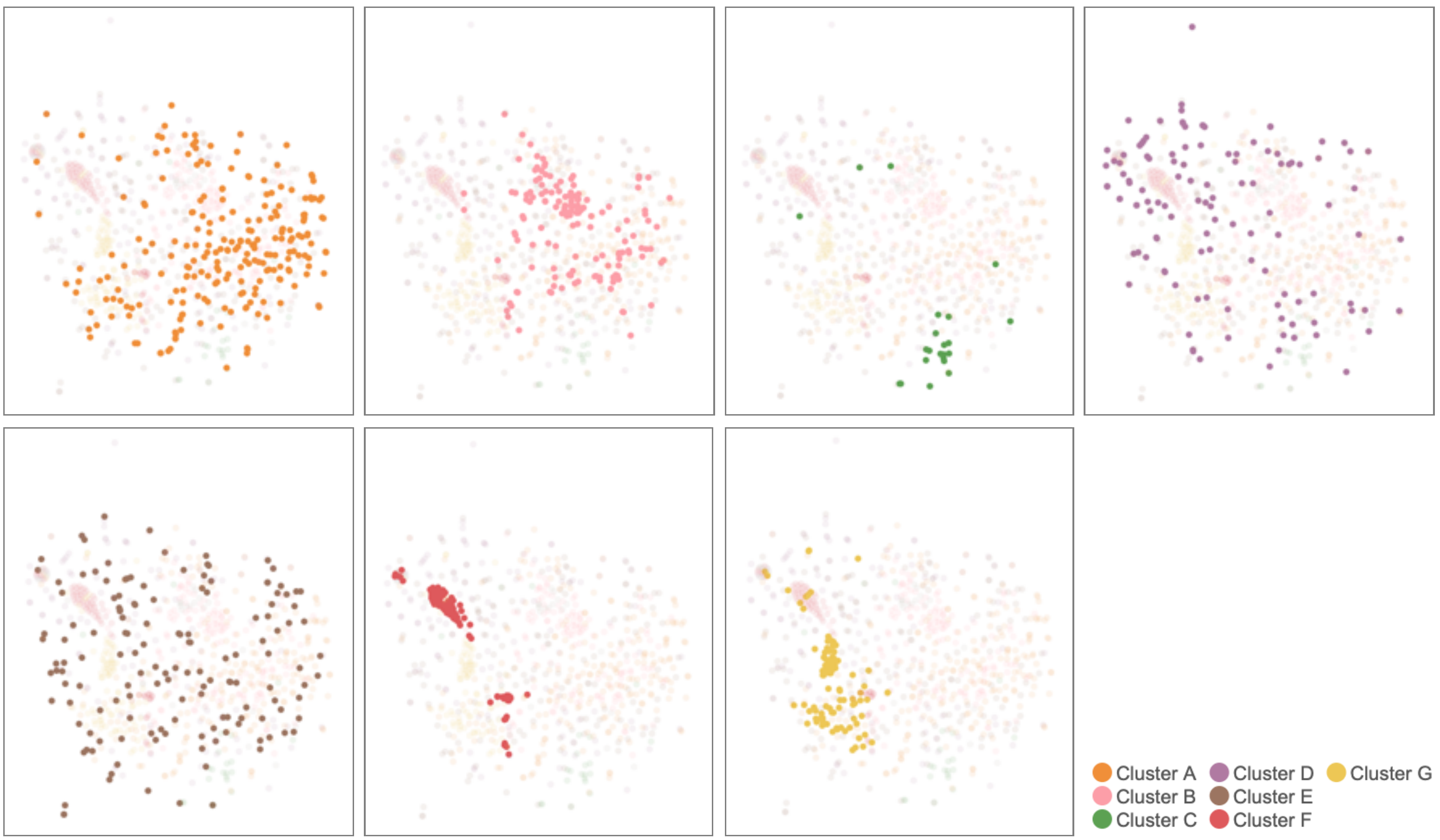}
    \caption{The similarities of neonates' test records shown in \autoref{fig:system_overview}b with highlighting of each cluster.}
    \label{fig:neonate_groups}
\end{figure}

We color each point based on the corresponding cluster assigned in \autoref{fig:system_overview}a to provide better linking between the views.
As shown in \autoref{fig:neonate_groups}, with these colors, we can see neonates tend to cluster together based on the test they took except for Cluster D (purple) and E (brown).

This view supports the same interactions implemented for \autoref{fig:system_overview}a, such as the neonate and clustering selections. 
Also, the views in \autoref{fig:system_overview}a and \autoref{fig:system_overview}b are fully linked with each other. 
For example, as shown in \autoref{fig:cluster_filtering}, when the selection is updated in the view of similarity of tests taken, the view of similarity of overall test records will be updated accordingly, and vice versa. 

\subsection{Transition of Test Records' Dissimilarities (\autoref{fig:system_overview}c)}

The transition of test records' dissimilarities in \autoref{fig:system_overview}c shows an overall (dis)similarity progression of the focal neonate and its top-3 similar records. 
The view helps users understand how the top-3 similar records are similar to the focal record, and when they are most or less similar (i.e., T2 in \autoref{sec:targets}).

To show the information above, we need to compute the similarity of these neonates' test results for each time step. 
One possible option is to use the vector representation after the sequence embedding in \autoref{sec:seq_embedding}. 
However, in this case, time steps with different lengths have been embedded into the same length of vectors. 
As a result, it is difficult to select a specific time step from the result, which is necessary to perform the ensuing tasks (i.e., T3 and T5).
Instead, we use the vector representation after the event embedding (i.e., the first-step embedding) in \autoref{sec:event_embedding}. 
However, because the selected neonates' records could have a different length of time steps, we cannot calculate the similarity of each time step directly. 
Thus, as similar to the work by Jin et al.~\cite{jin2019carepre}, which applies DTW~\cite{DTW} to the vectors after the event embedding, we first align the focal and top-3 similar records to the record with the longest time steps. 
Then, based on the alignment, we compute the similarity of the top-3 similar records' event vectors to the focal record for each time step.
Note that while Jin et al.~\cite{jin2019carepre} used DTW for computing similarities of records across all time steps, we use DTW only for the alignment of two vectors of different lengths for visual comparison.

To keep the visualizations simple, we use a line chart to show the computed transition of test records' dissimilarities, as shown in \autoref{fig:system_overview}c.
$x$- and $y$-coordinates represents a time step and the computed dissimilarity, respectively. 
Each polyline corresponding to one of the top-3 neonates is colored with the same scheme used for the outer-ring in \autoref{fig:system_overview}a and \autoref{fig:system_overview}b (e.g., the darkest blue shows the most similar neonate). 
A visualized example is shown in \autoref{fig:system_overview}c. 
In \autoref{fig:system_overview}c, we can see that all the top-3 neonates keep having relatively small dissimilarities across time steps (about 0.2 or less). 
This shows that our similarity calculation described in \autoref{sec:model} seems to properly identify similar neonate records.
On the other hand, we can also see a clear increase of dissimilarities after time step 7.

Also, while we show y-coordinates with a range of 0--1 as a default, the system allows the user to toggle zooming into a range from 0 to the maximum dissimilarity by clicking the `zoom' button placed at the top of the $y$-axis.
Lastly, the user can select a time step of interest by mouse-clicking. 
The selected time step is indicated with a black-dashed line (e.g., time step 9 in \autoref{fig:system_overview}c). 
As we describe in \autoref{sec:vis_at_selected_time}, the view in \autoref{fig:system_overview}e will be updated based on the selected time. 

\subsection{Overview of Test Records (\autoref{fig:system_overview}d)}
\label{sec:overview_test_record}

This view shows the statistical overview of test results of the focal and top-3 similar neonates. 
More specifically, for each neonate, we visualize the average value of each key test items across time steps.
32 key test items are selected out of 239 medical tests by filtering out the tests which were less frequently taken.
Specifically, we have picked the 8 most frequently-taken tests for each of the 7 clusters shown in \autoref{fig:system_overview}a.
Some of the test items are overlapped across clusters and, as a result, we obtain 42 test items in total. 
Then, we remove 10 test items, because they have near to or exactly zero standard deviations across neonates and time steps. 
This view is helpful for the user to find medical tests which tend to have (dis)similar results across the focal and top-3 similar neonate at a glance (T4 in \autoref{sec:targets}). 
To make it possible to compare across different medical tests, we have applied the standardization to each test item. 
With the standardization, all test items are converted to have zero mean and unit variance.
$y$-coordinates of this view represent the standardized value.
Similar to parallel-coordinates~\cite{inselberg1987parallel}, we use each vertical axis colored with light gray to show the values of each test item. 
Because of the space limitation, we only show the index of the test item at the bottom of each vertical axis. 
The corresponding test names are listed in the test items shown in \autoref{fig:system_overview}g.
Each polyline corresponds to one of the top-3 neonates with the same color scheme with \autoref{fig:system_overview}c. 
Also, a polyline for the focal neonate is colored with black as similar to the outer-rings in \autoref{fig:system_overview}a and \autoref{fig:system_overview}b.
As an interaction, the user can select a test item of interest by mouse-clicking. 
The selected test item will be indicated with the teal dashed-bar as shown in \autoref{fig:system_overview}d (`12: Lymphs'). 
Also, the related views (i.e., \autoref{fig:system_overview}e, f, and g) are also updated based on the selected item.

In a visualized example in \autoref{fig:system_overview}d, we can see that, generally, the focal and top-3 similar neonates' test results are similar to each other. 
However, three test items `10: Hematocrit', `11: Hemoglobin', and `25: Red cell count' have clear differences between the focal and others. 
The user may want to see more details of these two test items in the transition of these three items. 
The view of the transition of selected test's records described in \autoref{sec:vis_at_selected_item} can be used for this purpose. 

\subsection{Test Records at Selected Time (\autoref{fig:system_overview}e)}
\label{sec:vis_at_selected_time}

This view can be used to perform T4 in \autoref{sec:targets} and shows all medical test results for the focal and top-3 similar neonates at a selected time step in the view in \autoref{fig:system_overview}c or \autoref{fig:system_overview}f. 
The visualizations and interactions for this view are the same as \autoref{fig:system_overview}d. 
The only difference is that this view shows the actual test results instead of the average values across time steps. 

From the result in \autoref{fig:system_overview}e, we can understand the cause of the dissimilarity between the focal and other neonates at time step 9 in \autoref{fig:system_overview}c. 
The values for most tests of the top-3 similar records are similar to the focal neonate, whereas tests `10: Hematocrit', `11: Hemoglobin', `12: Lymphs', etc. have a large difference, especially `12: Lymphs'. 
The user may want to further investigate the overall transition patterns of these tests across time.
\autoref{fig:system_overview}f in the next subsection can be used for this analysis.

\subsection{Transition of Selected Test's Records (\autoref{fig:system_overview}f)}
\label{sec:vis_at_selected_item}

This view supports T5 in~\autoref{sec:targets} by providing a result to compare the neonates' detailed values of a certain test item across time. 
We employ visualizations and interactions similar to \autoref{fig:system_overview}c except for two differences.
On the $y$-axis, this view shows the standardized results, as explained in \autoref{sec:overview_test_record}, of the selected test item. 
A test name of the selected item is also shown as a $y$-axis label.
Additionally, because we want to compare the top-3 neonates' records with the focal neonate, this view includes the black polyline corresponding to the focal neonate record.

From the result in \autoref{fig:system_overview}f, we can see that while the focal neonate's `Lymphs' shows a  decrease at time step 8, right immediately after that, it radically increases to about 6. 
Because `Lymphs' is closely related to the immune system of the human body, the focal neonate seems to have some drastic changes in his/her immune system. 

\subsection{Linked Interactions across Multiple Views}
\label{sec:interaction}

As described, we have carefully designed visualizations and interactions to show linked information across views. 
Here we summarize all the linked interactions across multiple views.

\begin{asparadesc}
    \item[Selection of a focal neonate.] 
        The user can select a focal neonate by entering a neonate ID into the input dialog placed at the top left of \autoref{fig:system_overview}a or mouse-clicking a point displayed either in \autoref{fig:system_overview}a or \autoref{fig:system_overview}b. 
        After the selection, the system automatically selects the top-3 similar neonates' records and updates all views except for \autoref{fig:system_overview}g, accordingly. 
    \item[Selection of a cluster.]
        The user can select/unselect a cluster by mouse-clicking from the cluster legend placed in \autoref{fig:system_overview}a or \autoref{fig:system_overview}b. 
        This updates the top-3 similar neonates' records to be selected within the selected cluster. 
        Similar to the selection of a focal neonate, all views except for \autoref{fig:system_overview}g will be updated.
    \item[Selection of a time step.]
        The user can indicate a time step of interest in \autoref{fig:system_overview}c or \autoref{fig:system_overview}f by clicking one of $x$-coordinates. 
        This selection updates the black-dashed vertical lines in both \autoref{fig:system_overview}c and \autoref{fig:system_overview}f and the test item values shown in \autoref{fig:system_overview}e.
    \item[Selection of a test item.]
        A test item of interest can be selected in \autoref{fig:system_overview}d or \autoref{fig:system_overview}e by clicking one of the vertical axes.
        Also, this can be done with \autoref{fig:system_overview}g by clicking a test item.
        Afterward, the teal-dashed vertical lines in \autoref{fig:system_overview}d and \autoref{fig:system_overview}e and the rectangle with teal-stroke in \autoref{fig:system_overview}g will be updated. 
        Additionally, the system updates the visualized result in \autoref{fig:system_overview}f.
\end{asparadesc}

\section{Case Studies}

We demonstrate the effectiveness of our visual analytics system to identify and analyze similar neonates' records\footnote{An introduction video to the system regarding the case study can be found at \url{https://youtu.be/Ng89PDPcQpc}.}.

\subsection{Comparison of Similar Neonates Suffered from the Same Symptom}
We have already shown several findings by analyzing the records of ID 175 and similar patients while showing an example of usage of the system in \autoref{sec:vis}.
Here, we demonstrate a more comprehensive analysis of this neonate.

\begin{figure}[tb]
    \captionsetup{farskip=0pt}
	\centering
	\hspace*{-5pt}
	\subfloat[Average test values across time]{
     \includegraphics[width=0.49\linewidth]{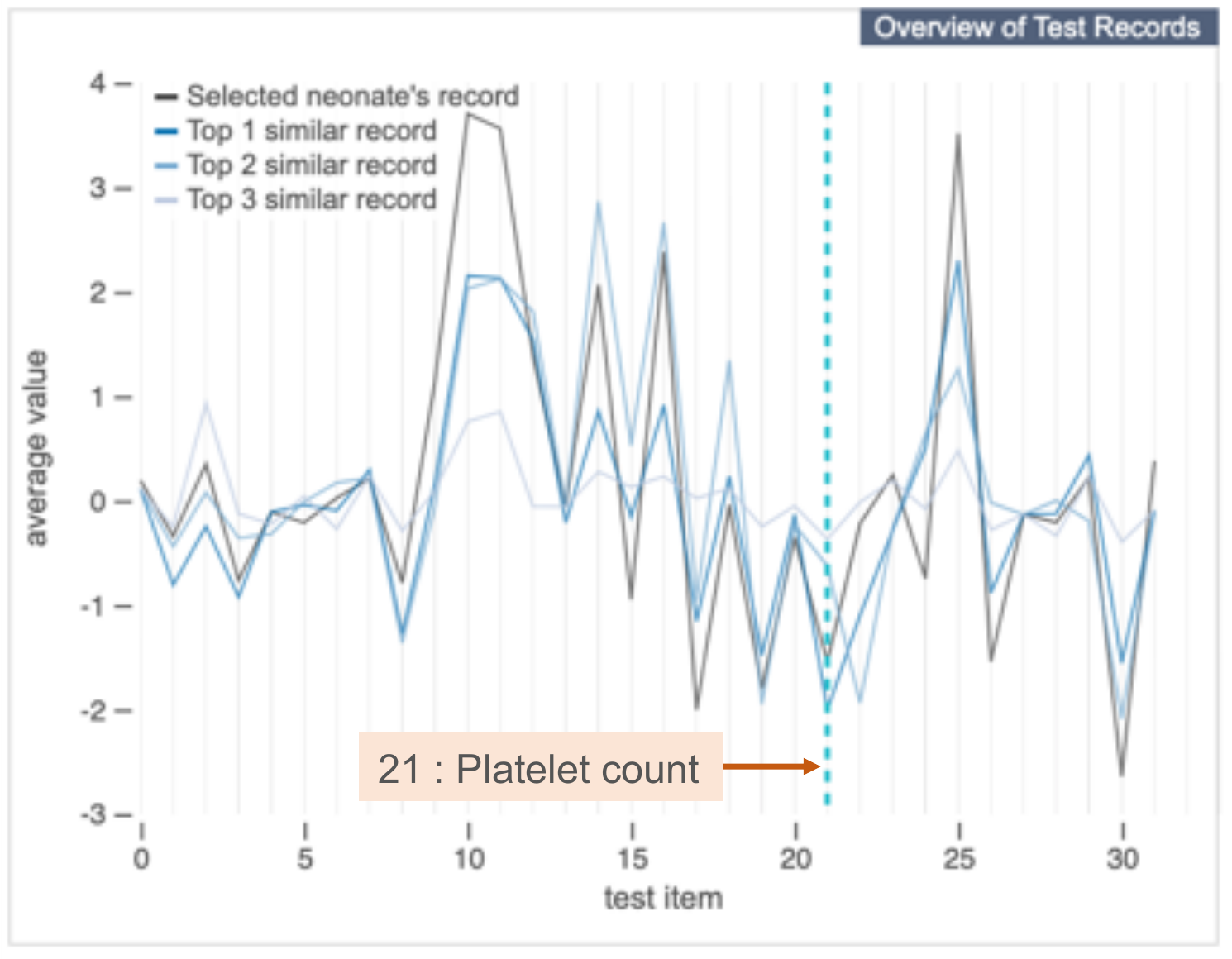}
     \label{fig:cs_1_1_a}
    }
    \hspace*{-3pt}
    \subfloat[Transition of platelet counts]{
     \includegraphics[width=0.49\linewidth]{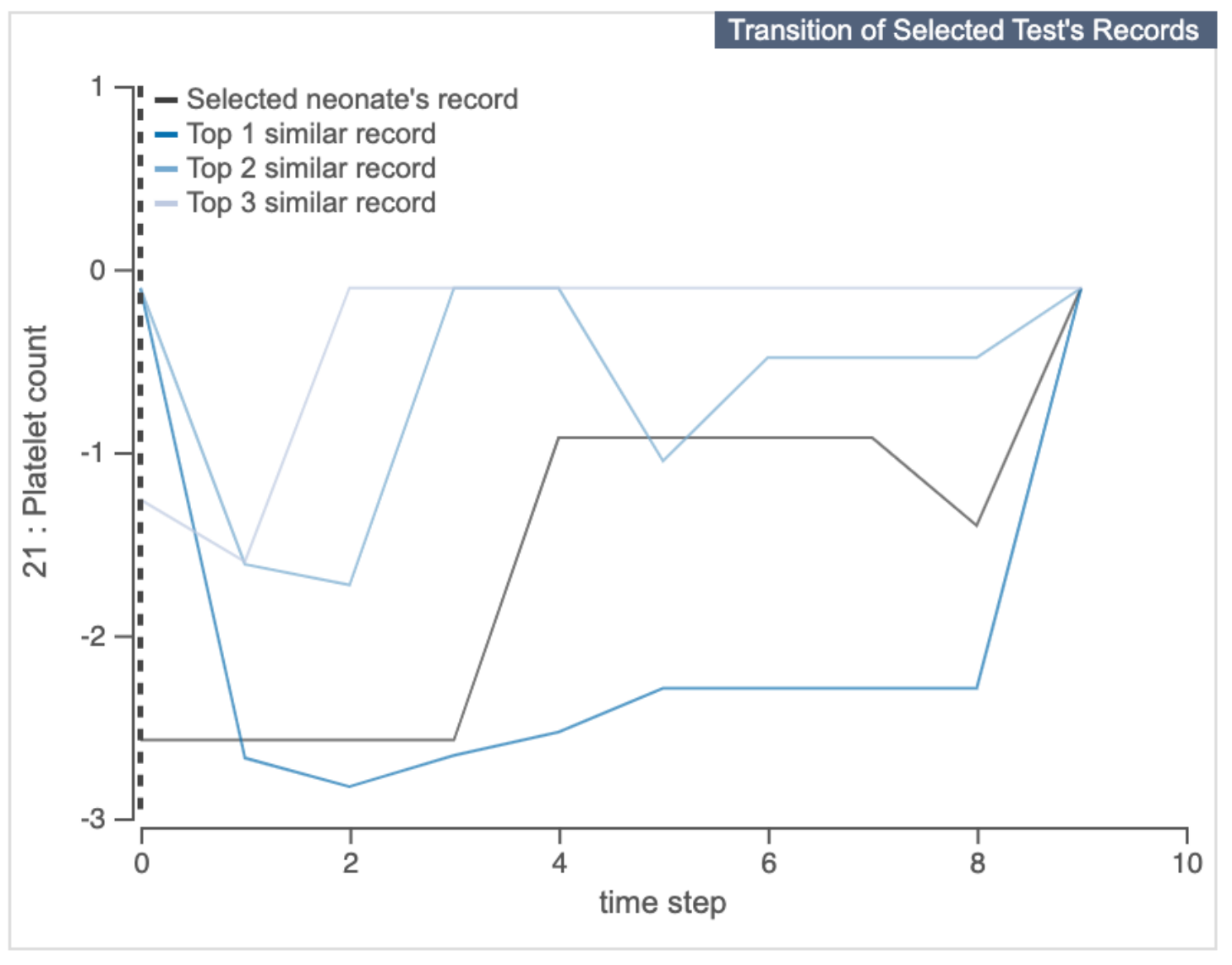}
     \label{fig:cs_1_1_b}
    }
   	\caption{The analysis of platelet counts of ID 175 and the top-3 similar neonates.}
	\label{fig:cs_1_1}
\end{figure}

Based on the documentation by a clinician in charge, ID 175 suffered from thrombocytopenia (i.e., the low platelet count in the blood). 
As shown in the overview of test records in  \autoref{fig:cs_1_1_a}, looking at values of `21: Platelet count' for these four neonates, we notice that the test values for all of them are lower than zero. 
Because we have applied standardization to each test item, this result indicates that these neonates have lower platelet counts than the average. 
First, we select `21: Platelet count' from this view and analyze the  detailed temporal changes of platelet counts in the view of the transition of selected records.
The result is shown in \autoref{fig:cs_1_1_b}.
We can see that all four neonates' platelet counts gradually increased and became the same amount as the average.
Thus, we can expect these neonates no longer suffer from thrombocytopenia. 

\begin{figure}[tb]
    \captionsetup{farskip=0pt}
	\centering
     \includegraphics[width=0.75\linewidth]{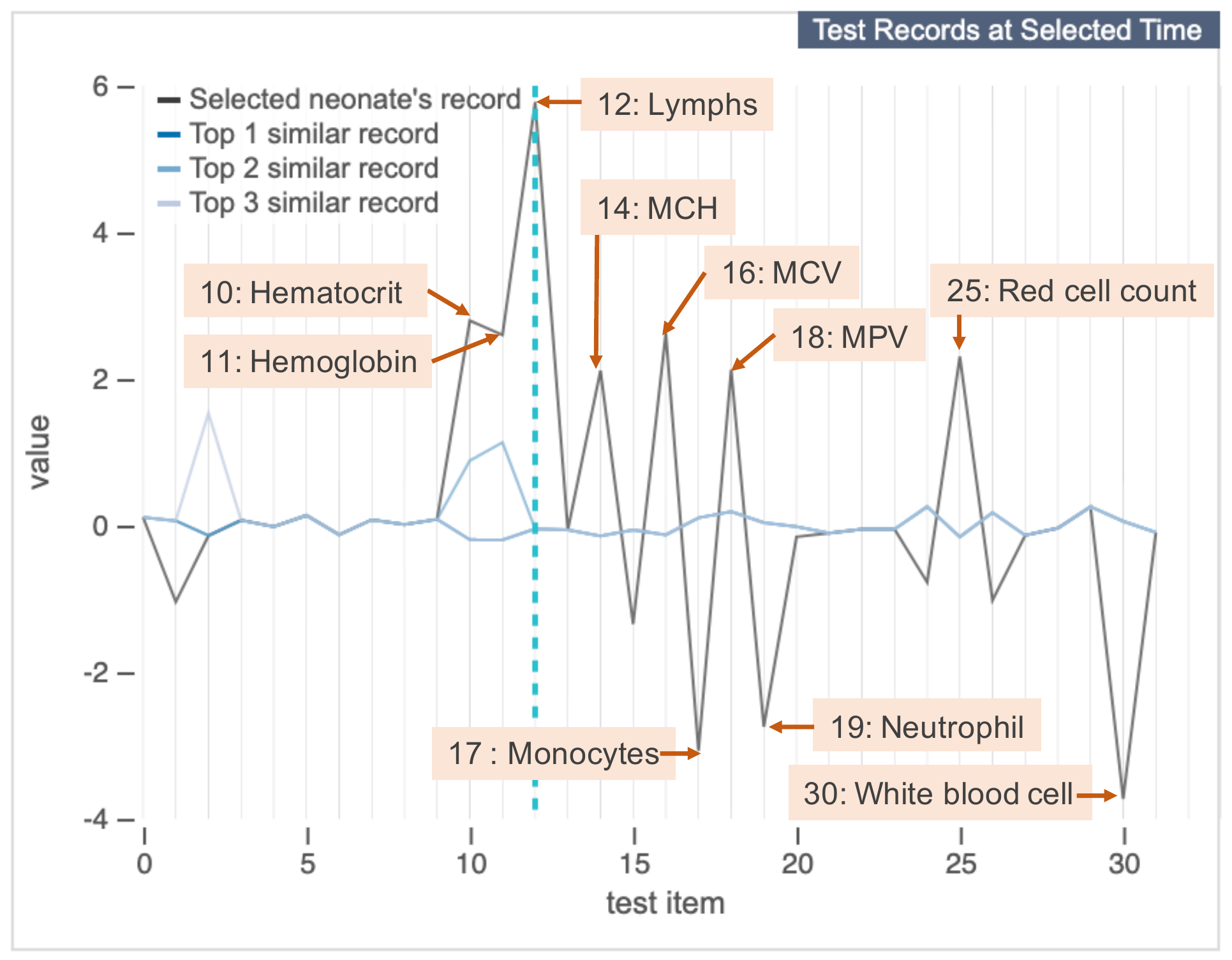}
   	\caption{The test values of ID 175 and the top-3 similar neonates at time step 9.}
	\label{fig:cs_1_2}
\end{figure}

\begin{figure}[tb]
    \captionsetup{farskip=0pt}
	\centering
	\hspace*{-5pt}
	\subfloat[30: White blood cell]{
     \includegraphics[width=0.49\linewidth]{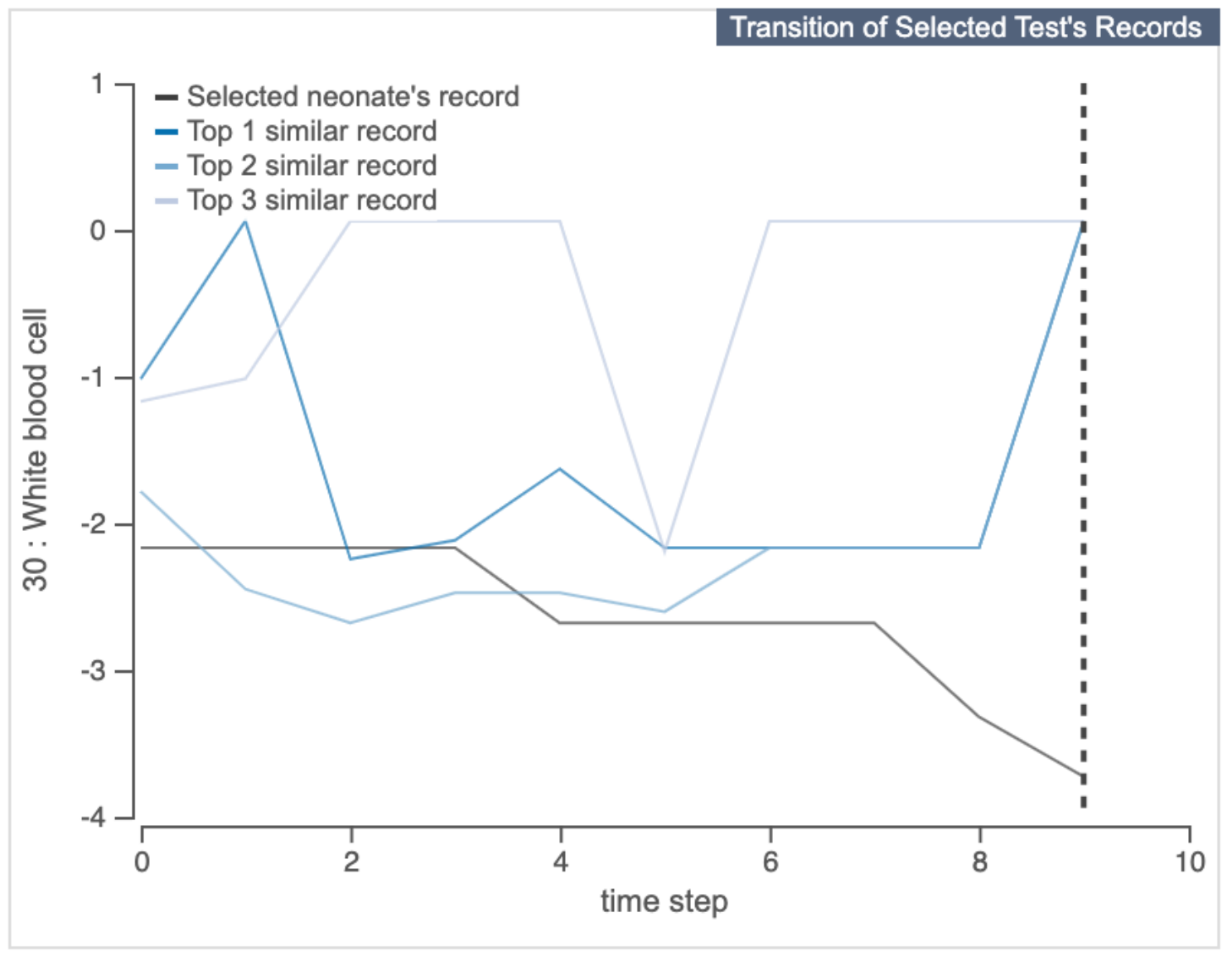}
     \label{fig:cs_1_3_a}
    }
    \hspace*{-5pt}
    \subfloat[11: Hemoglobin]{
     \includegraphics[width=0.49\linewidth]{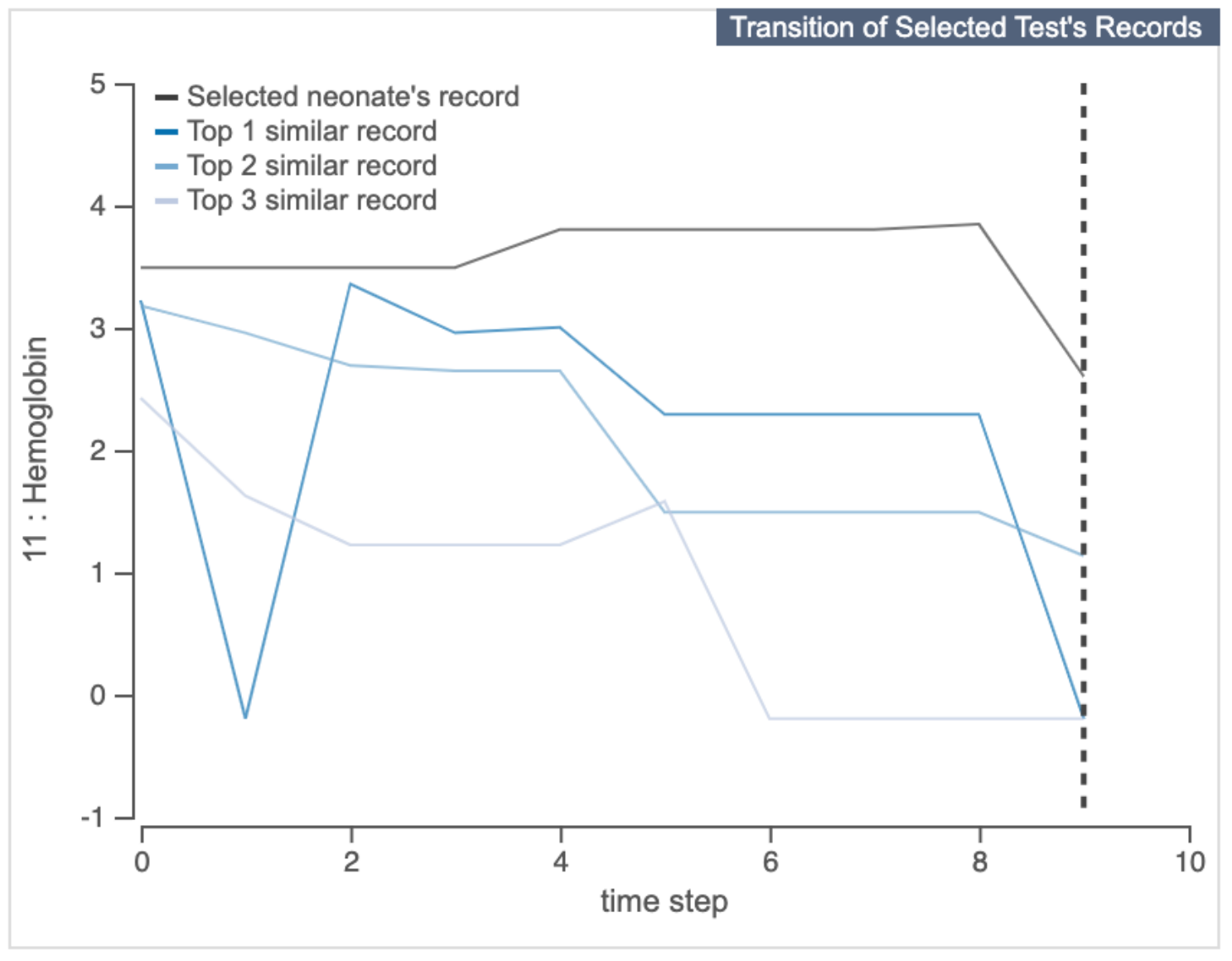}
     \label{fig:cs_1_3_b}
    }
    \\
    \hspace*{-5pt}
    \subfloat[25: Red cell count]{
     \includegraphics[width=0.49\linewidth]{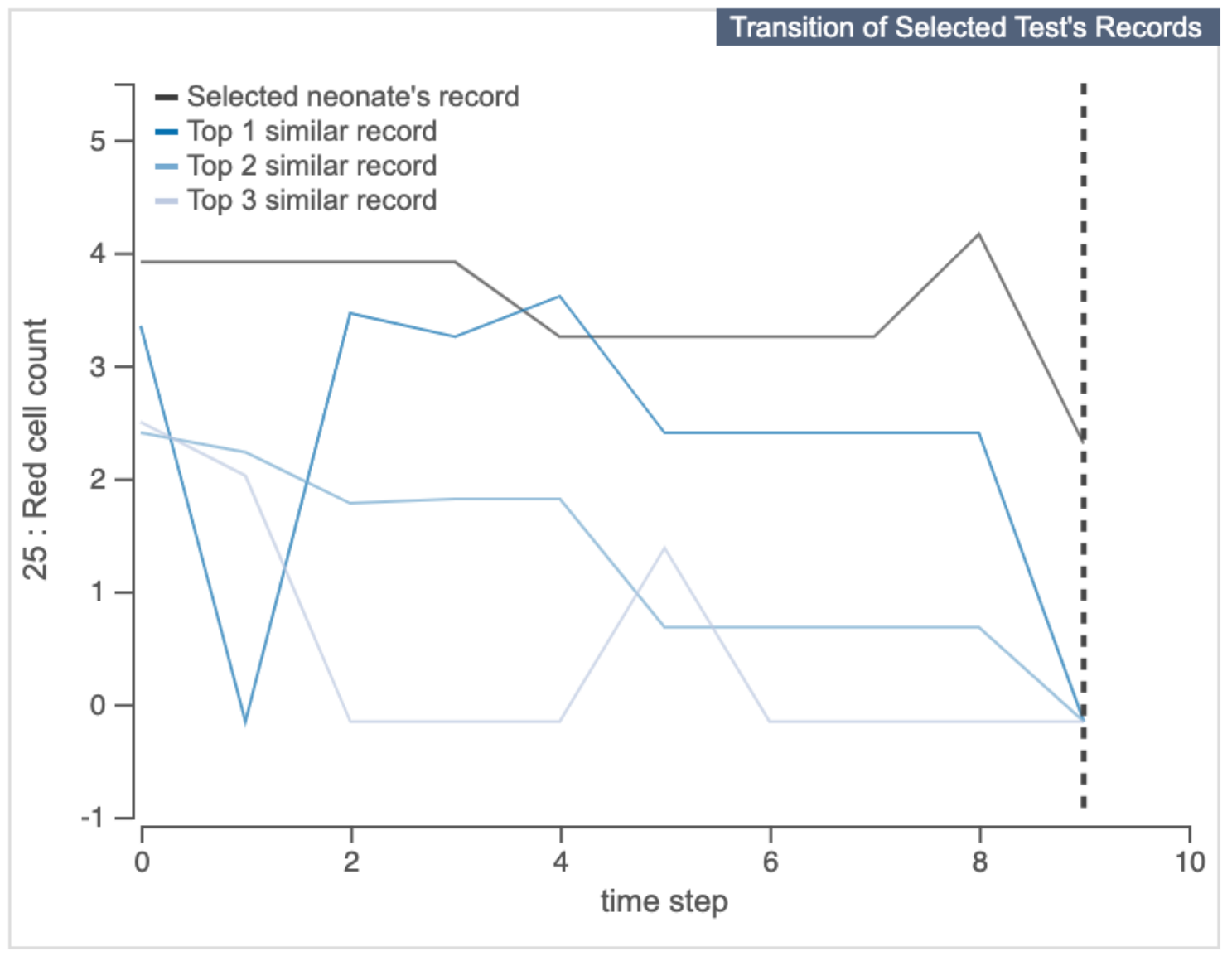}
     \label{fig:cs_1_3_c}
    }
    \hspace*{-5pt}
    \subfloat[18: MPV]{
     \includegraphics[width=0.49\linewidth]{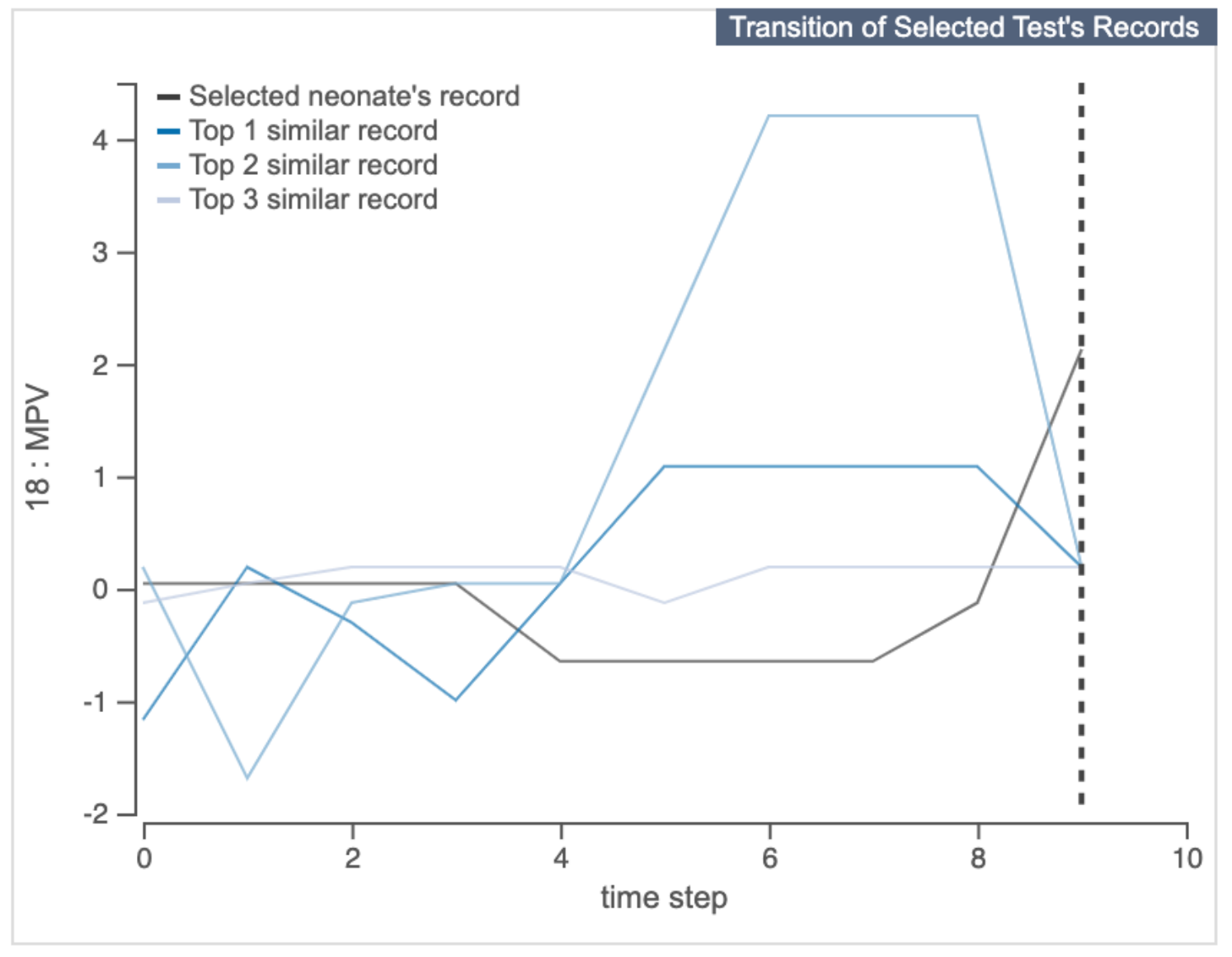}
     \label{fig:cs_1_3_d}
    }
   	\caption{The details of temporal changes of test items that take small (`30: White blood cell') or large (`11: Hemoglobin', `25: Red cell count', and `18: MPV') values in \autoref{fig:cs_1_2}.}
	\label{fig:cs_1_3}
\end{figure}

However, as described in \autoref{sec:vis}, we remind that ID 175 has an increase of dissimilarity from the others at time step 9 and high `12: Lymphs' at that time step. 
We review the same plot, as shown in \autoref{fig:cs_1_2}.
We can see that, including `12: Lymphs', ID 175 has high values in several tests, such as `10: Hematocrit', `11: Hemoglobin', `14: MCH' (mean corpuscular hemoglobin), `16: MCV' (mean corpuscular volume), `18: MPV' (mean platelet volume), and `25: Red cell count'. 
On the other hand, `17: Monocytes', `19: Neutrophil', and `30: White blood cell' show low values.
Note that both `17: Monocytes' and `19: Neutrophil' are specific types of white blood cells.

We review temporal changes in all of these test items with the view of the transition of selected test's records and \autoref{fig:cs_1_3} shows some of these results.
From the values of `30 : White blood cell' shown in \autoref{fig:cs_1_3_a}, we can see that the total number of white blood cells of ID 175 keeps going far from the average value.
Also, from \autoref{fig:cs_1_3_b} and \autoref{fig:cs_1_3_c}, we can see that ID 175 keeps high values for `11: Hemoglobin' and `25: Red cell count' while other neonates has about the average (i.e., zero value) at time step 9. 
This indicates that ID 175 still suffered from polycythemia at his last hospital visit.
Also, more importantly, in \autoref{fig:cs_1_3_d}, we can see an increase of MPV in ID 175, which indicates the average size of the platelet becomes large. 
High MPV implies a potential cause of thrombocytopenia~\cite{arad1986mean}. 
Therefore, unlike the other similar neonates, even though we observed improvement in `21: Platelet count', as shown in \autoref{fig:cs_1_1_b}, ID 175 should be kept checking his/her blood. 

\subsection{Analysis of Neonate Groups}
We perform an analysis of similar neonates selected within a cluster. 
As we have shown in \autoref{fig:cluster_filtering}, several distinct clusters of neonates can be observed in the view of similarity of overall test records (\autoref{sec:similarity_test_records}). 
In this case study, we analyze two distinct clusters, Cluster C (green) and F (red).

\begin{figure}[tb]
    \captionsetup{farskip=0pt}
	\centering
	\hspace*{-5pt}
	\subfloat[Similarities of overall test records]{
     \includegraphics[width=0.49\linewidth]{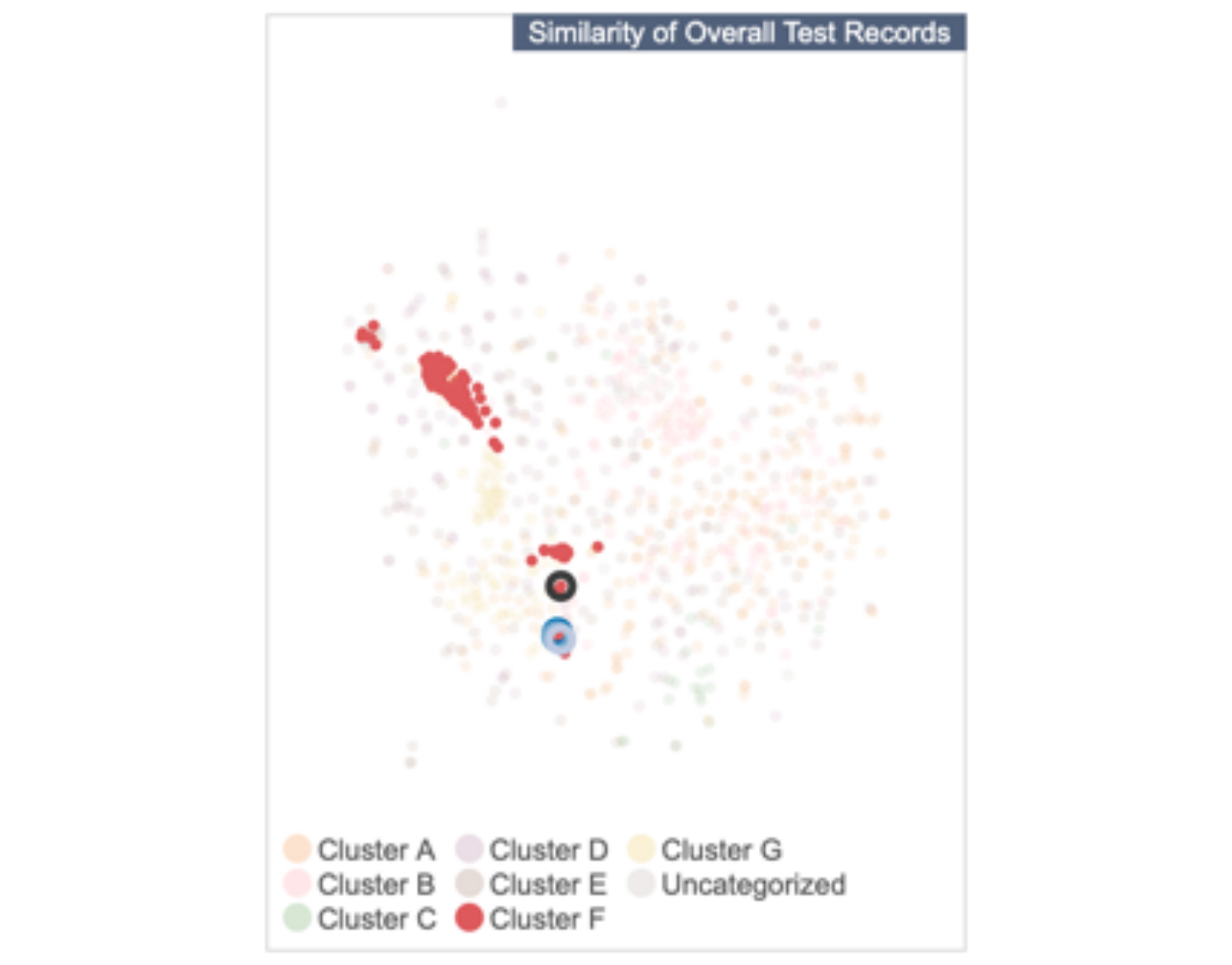}
     \label{fig:cs_2_1_a}
    }
    \hspace*{-5pt}
    \subfloat[Transition of records' dissimilarity]{
     \includegraphics[width=0.49\linewidth]{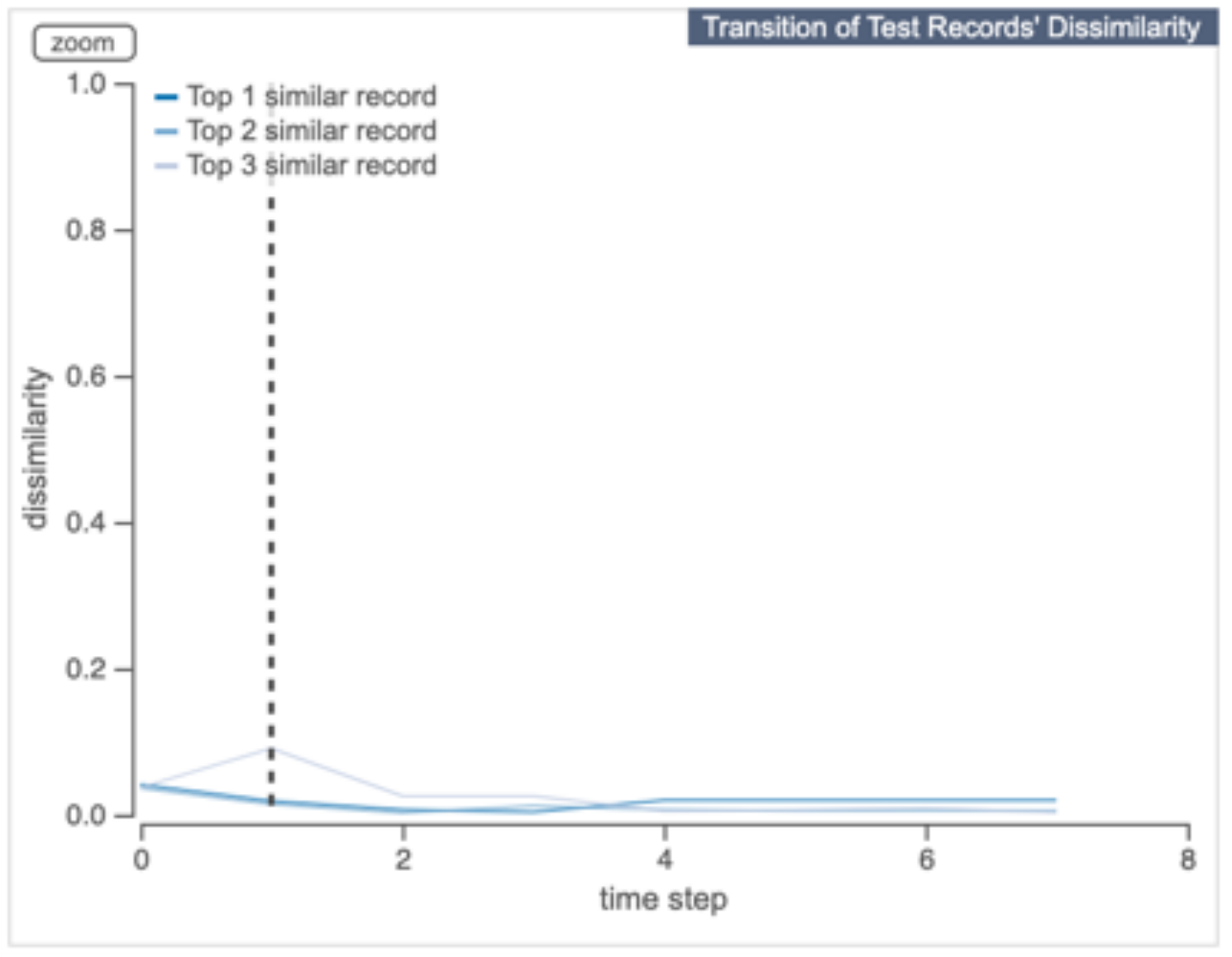}
     \label{fig:cs_2_1_b}
    }
    \\
    \hspace*{-5pt}
    \subfloat[Overview of values of all test items]{
     \includegraphics[width=0.49\linewidth]{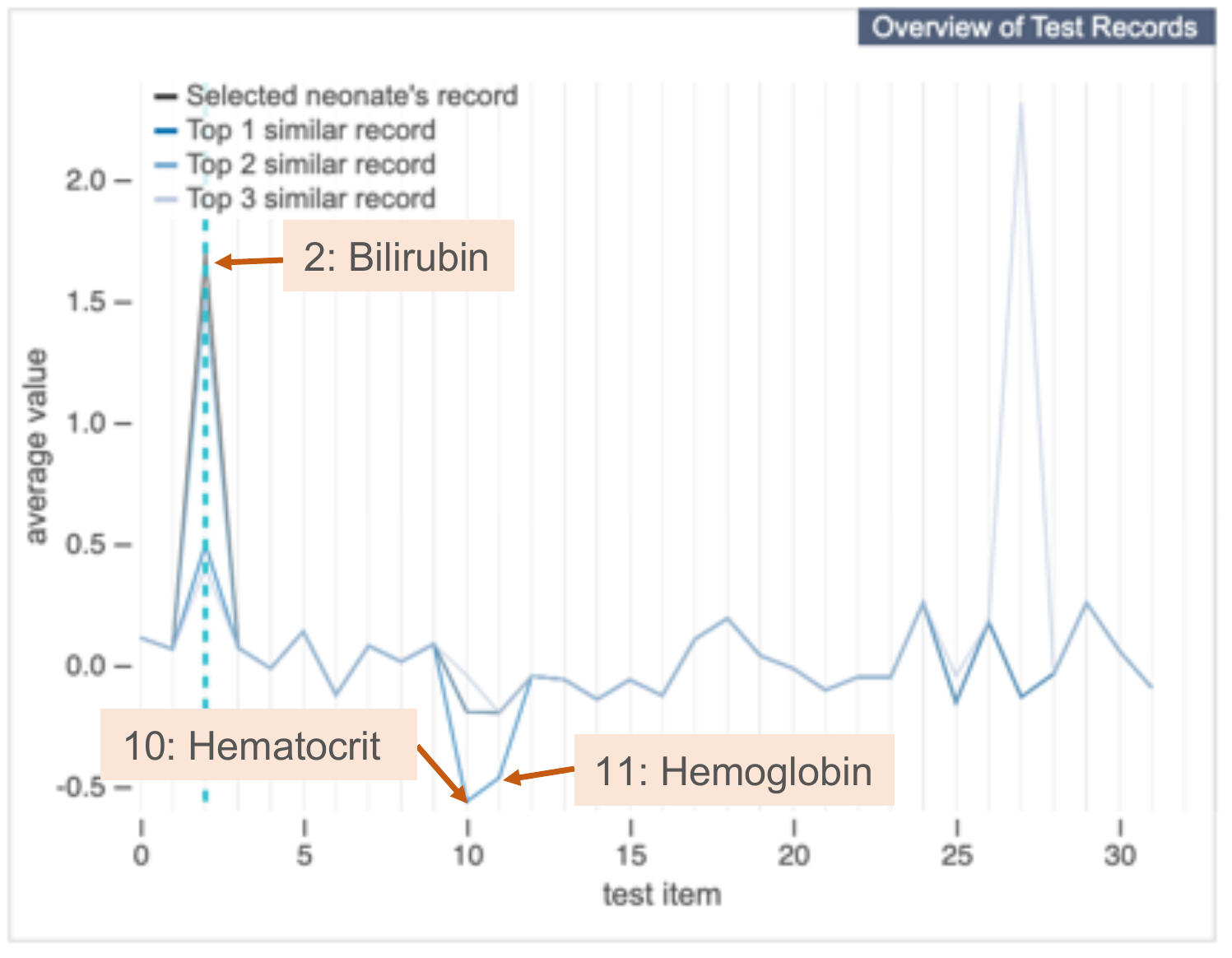}
     \label{fig:cs_2_1_c}
    }
    \hspace*{-5pt}
    \subfloat[Transition of `2: Bilirubin total']{
     \includegraphics[width=0.49\linewidth]{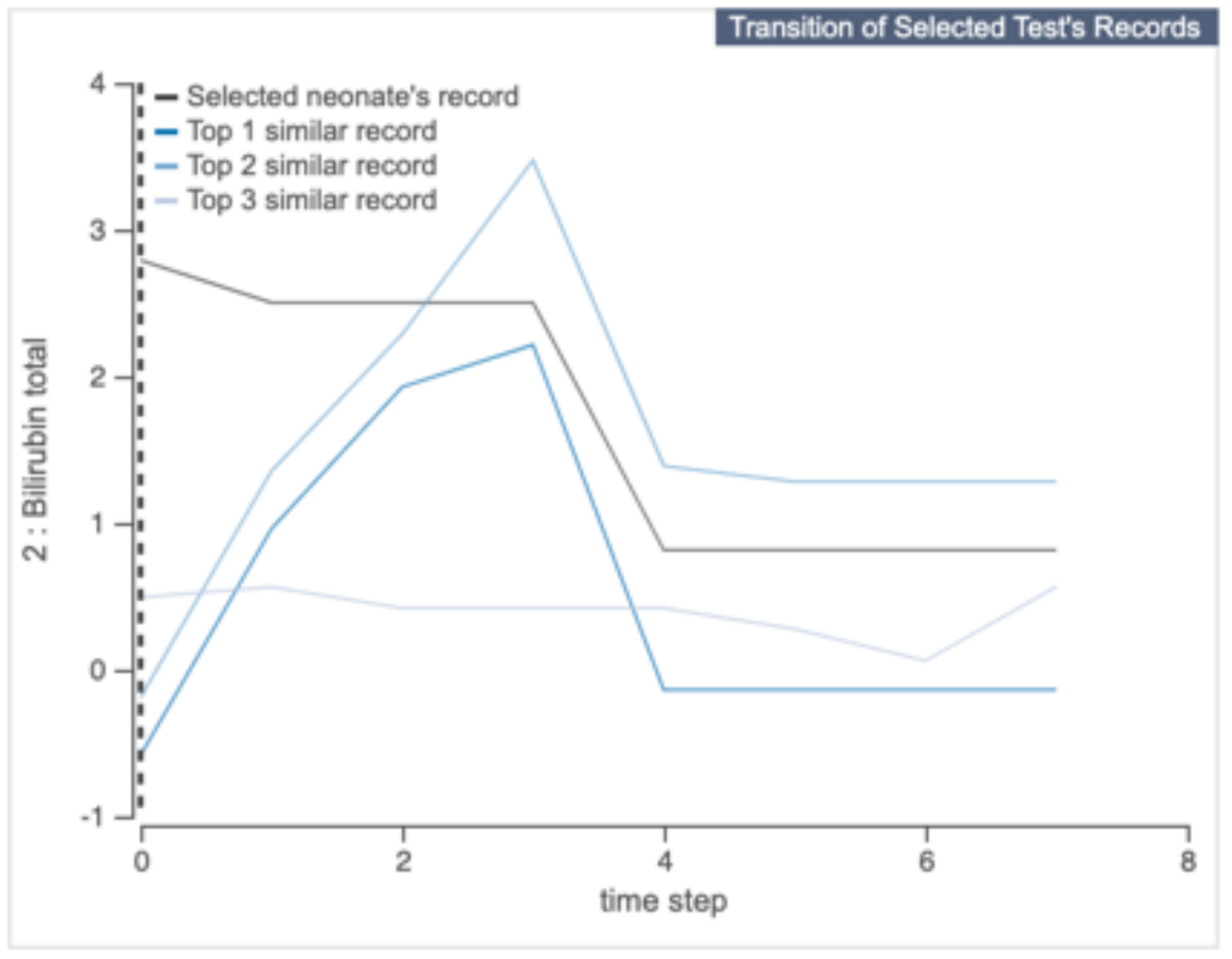}
     \label{fig:cs_2_1_d}
    }
   	\caption{The analysis of neonates in Cluster F.}
	\label{fig:cs_2_1}
\end{figure}

First, with the interaction of selection of a cluster (refer to \autoref{sec:interaction}), we select the focal and top-3 similar patients within Cluster F, as shown in \autoref{fig:cs_2_1_a}. 
Then, the related views are immediately updated. 
The results of the transition of test records' dissimilarities and overview of values of all test items are visualized in \autoref{fig:cs_2_1_b} and \autoref{fig:cs_2_1_c}. 
From \autoref{fig:cs_2_1_b}, we can see these four neonates keep high similarity with each other across time steps. 
Also, from \autoref{fig:cs_2_1_c}, we can see the neonates tend to have high `2: Bilirubin total'; low `10: Hematocrit' and `11: Hemoglobin'. 
This implies that the patients in Cluster F seem to be suffering from the destruction of hemoglobin and resultant high bilirubin.
When we review the transition of `2: Bilirubin total', as shown in \autoref{fig:cs_2_1_d}, most of the selected neonates tend to have high bilirubin in the early or middle stage and then show a decrease to a lower value as signs of recovery.

\begin{figure}[tb]
    \captionsetup{farskip=0pt}
	\centering
	\hspace*{-5pt}
	\subfloat[Similarities of overall test Records]{
     \includegraphics[width=0.49\linewidth]{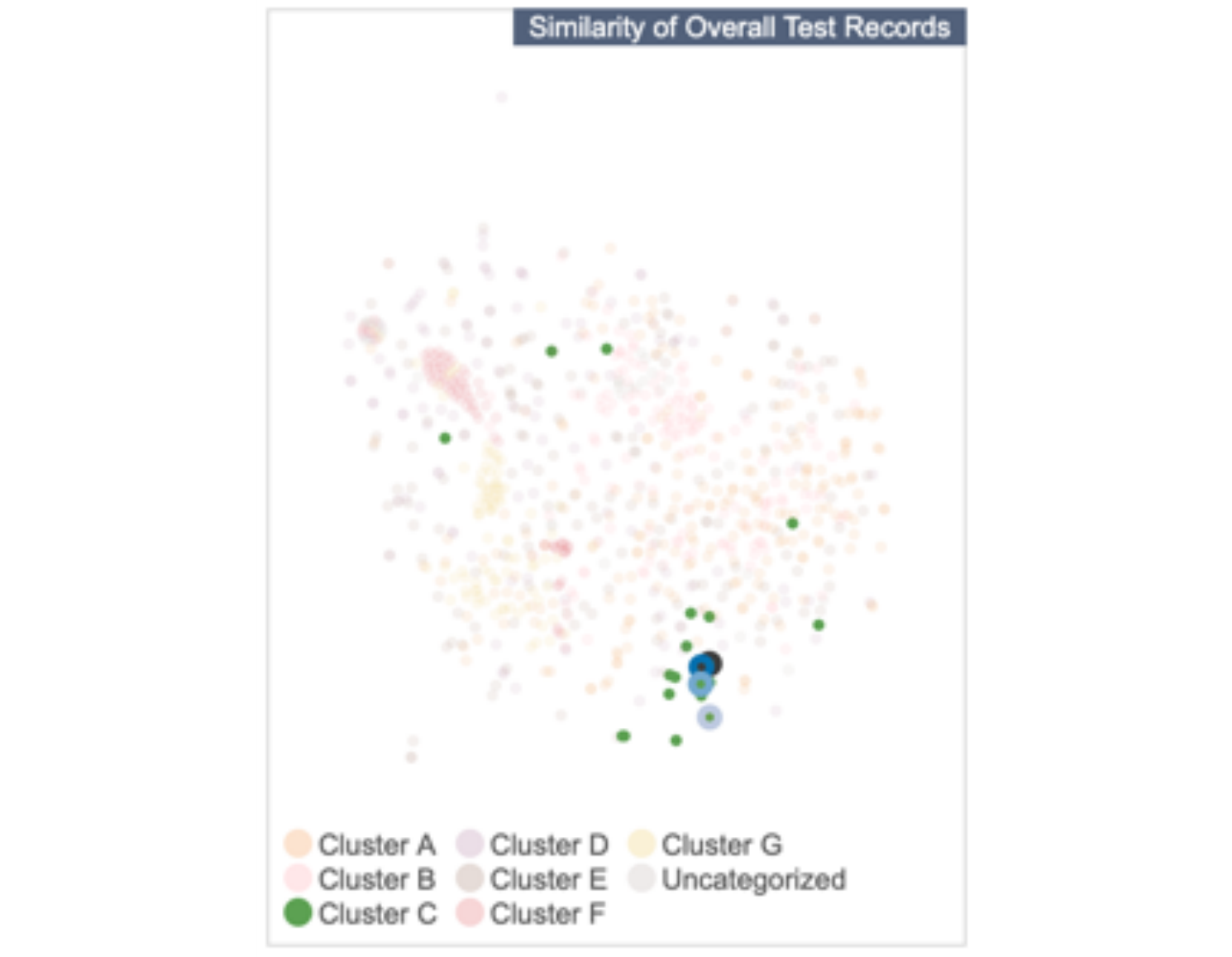}
     \label{fig:cs_2_2_a}
    }
    \hspace*{-5pt}
    \subfloat[Transition of records dissimilarity]{
     \includegraphics[width=0.49\linewidth]{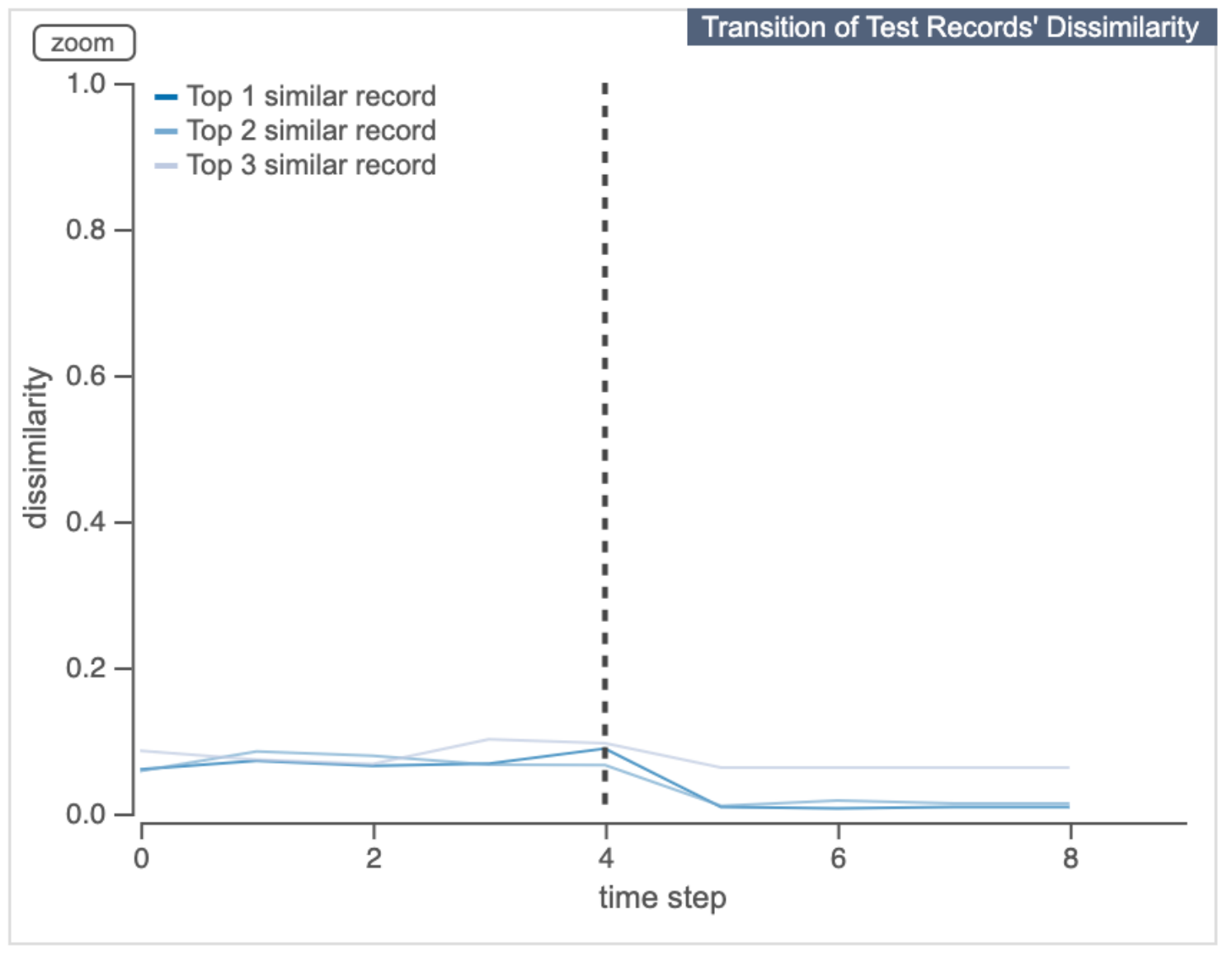}
     \label{fig:cs_2_2_b}
    }
    \\
    \hspace*{-5pt}
    \subfloat[Overview of values of all test items]{
     \includegraphics[width=0.49\linewidth]{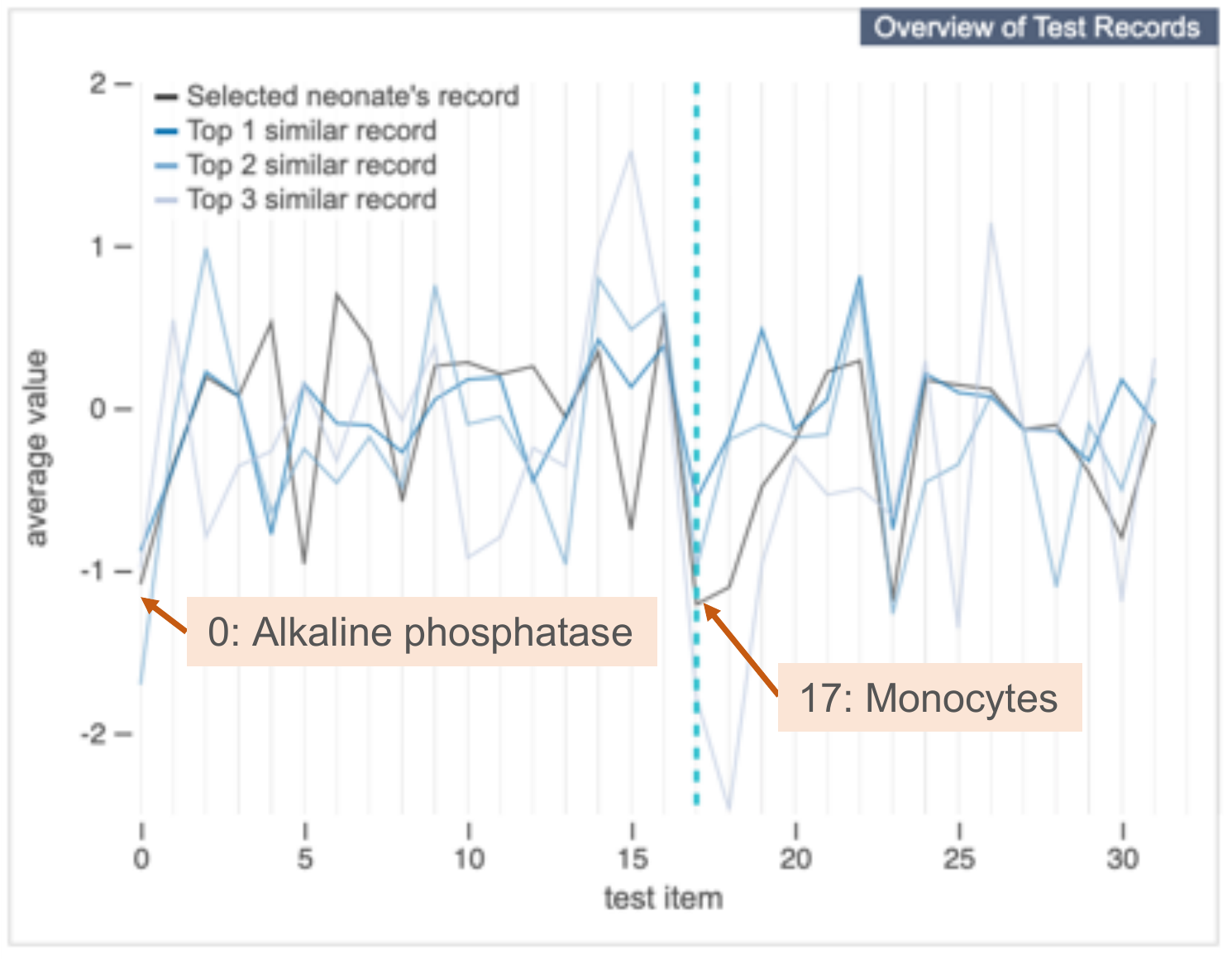}
     \label{fig:cs_2_2_c}
    }
    \hspace*{-5pt}
    \subfloat[Transition of `17: Monocytes']{
     \includegraphics[width=0.49\linewidth]{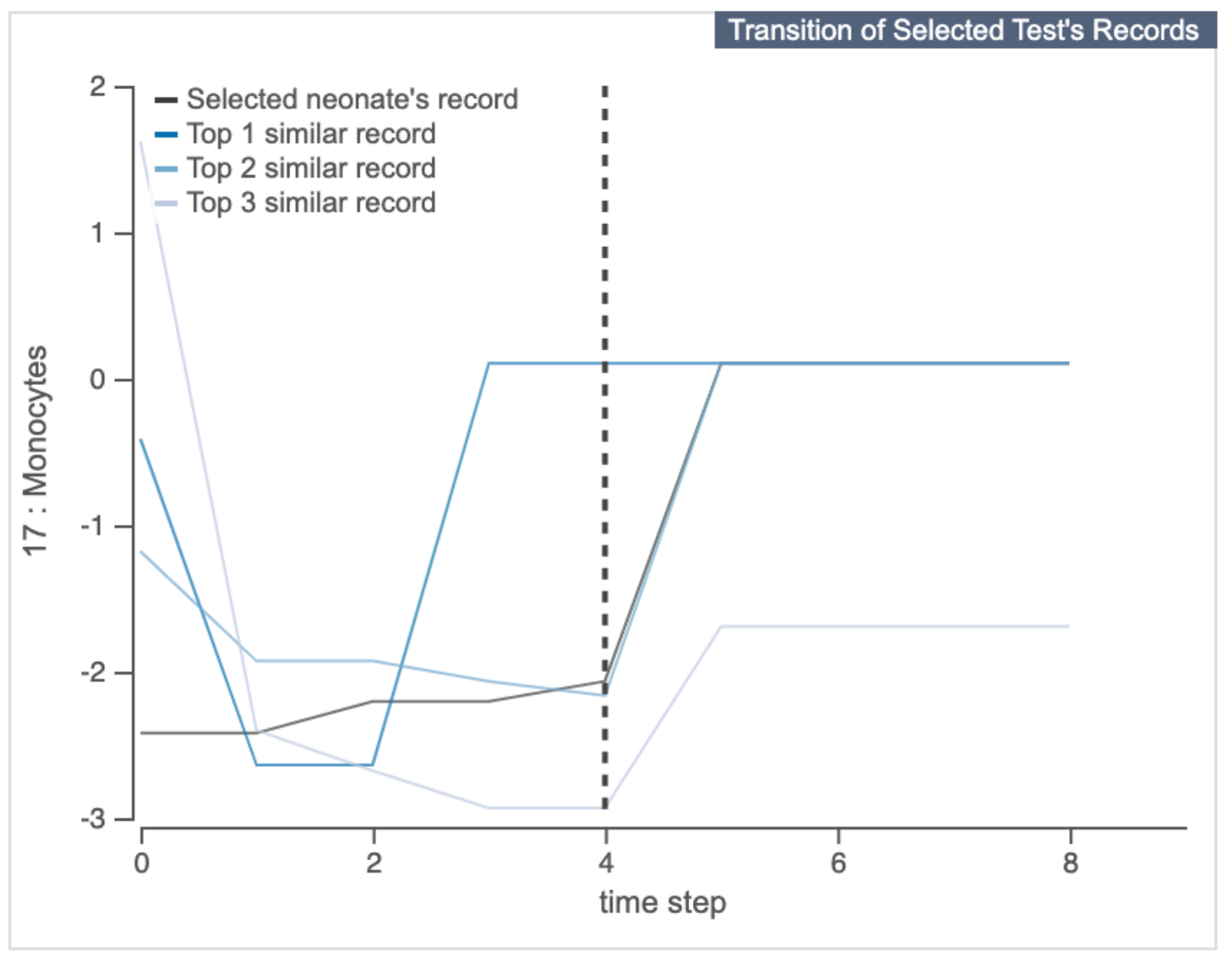}
     \label{fig:cs_2_2_d}
    }
   	\caption{The analysis of neonates in Cluster C.}
	\label{fig:cs_2_2}
\end{figure}

Next, similarly, we select the neonates from Cluster C, as shown in \autoref{fig:cs_2_2_a}. 
From \autoref{fig:cs_2_2_b}, which depicts the transition of dissimilarities, we can see that these neonates also keep taking high similarities across time.
In the overview of values of all test items shown in \autoref{fig:cs_2_2_c}, we can see that the neonates have salient values in several test items.
For example, `0: Alkaline phosphatase' and `17: Monocytes' have much lower values than the average of all neonates (i.e., zero value).
\autoref{fig:cs_2_2_d} shows the transition of values of `17: Monocytes'. 
From this, we can see that the neonates begin to take a close value to the average of all neonates while they used to suffer from low monocytes. 

This case study demonstrates the functionality of our system to analyze and compare groups of patients clustered by their symptoms by using the cluster selection.
\section{Discussion and Limitations}

\noindent\textbf{Extensive Usage of Our Embedding Method}
We have developed our two-step embedding described in \autoref{sec:model} to identify and compare similar medical records. 
The fundamental contribution of our embedding is enabling the similarity calculation of event sequences with different lengths. 
Similarities obtained from event sequences can be used not only for the identification of similar event sequences but also for many other data mining tasks, such as clustering and classification.
Therefore, we plan to extend the usage of our embedding for these types of analyses.

\vspace{5pt}
\noindent\textbf{Generality of our methods.}
We use our embedding method and visual analytics system to analyze neonates' medical test results.
However, our methods are generic enough to apply to other medical records that can be represented as multivariate time sequences without any major modifications. 
Moreover, we can use our methods for other applications, such as career decision support, financial analysis, and cybersecurity systems.
For instance, as similar to the existing works~\cite{du2016eventaction,du2017finding,du2019visual}, in a career decision support system, we can suggest similar students or workers to help the user select the next career. 

\vspace{5pt}
\noindent\textbf{Limitations of Our Embedding Method.}
Our embedding method is designed for event sequence data in which each event contains multivariate values. 
Thus, our method is not suitable for directly applying to event sequences that do not have any value at each event (e.g., each event only has an event name, such as a `hospital visit').
For such data, we need to modify the first step of embedding (i.e., the event embedding). 
For example, we can use the skip-gram~\cite{mikolov2013efficient} based embedding as used in the work by Guo et al.~\cite{guo2018visual}, or the CBOW~\cite{mikolov2013efficient} based embedding.
A more challenging situation is that we need to handle both types of event sequences above.
That is, some events contain multivariate values while some have only an event name. 
We would like to address this challenge as our future work. 
For example, we can assign a unique event name for an event with multivariate values based on the range of each value (e.g., high blood pressure and low heartbeat) and then apply the skip-gram or CBOW~\cite{mikolov2013efficient} based embedding.

\vspace{5pt}
\noindent\textbf{Limitations of Our Dataset and Analysis}
Through the paper, we use the neonate medical records consisting of only values for each medical test item.
Because this dataset does not include any medical judgements as events and we did not fully incorporate medical documentations (e.g., clinicians' comments), the insights obtained through the performed analyses would be limited and also need additional evaluations.
For example, by coupling with the clinician's judgement for each time step, we can correlate the transitions of test values and the diagnoses for the neonates to gain more detailed medical insights.   
Also, we would like to further understand clinical rationale in the cluster formation of similar neonate records, which is observed in \autoref{fig:neonate_groups}, with the expert knowledge.

\vspace{5pt}
\noindent\textbf{Limitations of Our Visual Analytics System.}
Our visual analytics system focuses on effectively supporting exploration of similar medical records from both temporal and multivariate perspectives.
However, the system does not provide an interface to investigate raw events, including clinicians' comments and the detailed date and time of tests taken.
We plan to integrate such functionalities into the sytem in the future.

We have chosen to use simple charts (scatterplots and line charts) to provide easy-to-understand visualization in our visual analytics system.
However, both scatterplots and line charts would have a scalability problem when we need to visualize more information.
For example, scatterplots are used to show the similarities of all 854 neonates records. 
However, some medical records contain a much larger amount of patients, for example, over 40,000 patients in the MIMIC-III dataset~\cite{mimiciii}, an open-access clinical database.
To handle such a large dataset, we should couple with data reduction methods. 
For example, we can use the data sampling method developed for scatteplots~\cite{hu2019data}.
As for line charts, we use them for showing the focal and top-3 similar neonates. 
When the user wants to visualize more lines, for example, showing the top-10 similar neonates, visualized results could suffer from cluttering of lines and/or difficulty in distinguishing the line colors.
For example, we have used the colors with the same hue but different saturation to avoid using the same hue with the cluster colors.
However, we can expect that this color scheme would work only when we visualize less than or equal to 5 similar neonates. 
When visualizing more lines is needed, we can consider using aggregation. 
For example, based on the similarity of each line, we can aggregate multiple similar lines in one line.
\section{Conclusions}
Identifying and analyzing similar patients' medical records are fundamental needs in clinical decision making. 
Our work provides an unsupervised learning-based method for measuring the similarity of medical records, which can deal with high-dimensionality, irregularity, and sparsity in medical records.
The visual analytics system built on top of our methods enables effective analysis of medical records from both temporal and multivariate perspectives.
In the future, we would like to conduct user studies with clinical researchers to understand more preferable visualization designs for them.
With the contributions above, we believe that our work better guides potential directions of future research on the comparative analysis of event sequence data in various domains.

\section*{Acknowledgments}
The authors wish to thank Dr.~Mark A.~Underwood at UC Davis Children's Hospital. This research is sponsored in part by the U.S. National Science Foundation through grant IIS-1741536 and a 2019 Seed Fund Award from CITRIS and the Banatao Institute at the University of California.

\bibliographystyle{abbrv-doi}
\bibliography{00_main.bib}

\end{document}